\pdfoutput=1

\documentclass[a4paper,cits]{JINST}
\usepackage{lineno}
\usepackage{microtype}
\usepackage{booktabs}
\usepackage{amsmath}
\usepackage[hyphens]{url}

\newcommand{\Eqe}{\ensuremath{\mathrm{E_{QE}}}}
\newcommand{\Enu}{\ensuremath{\mathrm{E_{\nu}}}}
\newcommand{\Ne}{\ensuremath{\mathrm{N_{e}}}}

\newcommand{\Ngamma}{\ensuremath{\mathrm{N_{\gamma}}}}
\newcommand{\Edepo}{\ensuremath{\mathrm{E_{depo}}}}
\newcommand{\Nq}{\ensuremath{\mathrm{N_{q}}}}
\newcommand{\Emiss}{\ensuremath{\mathrm{E_{miss}}}}
\newcommand{\Emissmu}{\ensuremath{\mathrm{E_{miss,\mu}}}}
\newcommand{\orderof}{\ensuremath{\mathcal{O}}}

\title{Expected performance of an ideal liquid argon neutrino detector with enhanced sensitivity to scintillation light}

\author{
M.~Sorel \\
Instituto de F\'isica Corpuscular (IFIC), CSIC \& Universitat de Val\`encia\\
Calle Catedr\'atico Jos\'e Beltr\'an, 2, 46980 Paterna, Valencia, Spain\\

E-mail: \email{sorel@ific.uv.es}
}

\abstract{
Scintillation light is used in liquid argon (LAr) neutrino detectors to provide a trigger signal, veto information against cosmic rays, and absolute event timing. In this work, we discuss additional opportunities offered by detectors with enhanced sensitivity to scintillation light, that is with light collection efficiencies of about $10^{-3}$. We focus on two key detector performance indicators for neutrino oscillation physics: calorimetric neutrino energy reconstruction and neutrino/antineutrino separation in a non-magnetized detector. Our results are based on detailed simulations, with neutrino interactions modelled according to the GENIE event generator, while the charge and light responses of a large LAr ideal detector are described by the Geant4 and NEST simulation tools. A neutrino energy resolution as good as 3.3\% RMS for 4 GeV electron neutrino charged-current interactions can in principle be obtained in a large detector of this type, by using both charge and light information. By exploiting muon capture in argon and scintillation light information to veto muon decay electrons, we also obtain muon neutrino identification efficiencies of about 50\%, and muon antineutrino misidentification rates at the few percent level, for few-GeV neutrino interactions that are fully contained. We argue that the construction of large LAr detectors with sufficiently high light collection efficiencies is in principle possible.
}


\begin{document}


\tableofcontents

\section{Introduction} \label{sec:Introduction}

Liquid argon time projection chambers (LAr TPCs) have first been proposed as neutrino detectors in 1977 \cite{Rubbia:1977zz}. In such detectors, charged particles produced by neutrino interactions deposit energy by ionizing and exciting argon. Ionization electrons can be collected by establishing a drift electric field between cathode and readout planes. The full three-dimensional image of the ionization signal in the active volume can be reconstructed from the two-dimensional image at the readout planes as a function of electron drift time, following the TPC detection principle. Liquid argon offers several advantages as neutrino detection medium. It is cheap, easy to obtain and to purify. It has a relatively high boiling point, so that it can be liquefied with liquid nitrogen. In its liquefied form, argon has a high density of 1.39 g/cm$^3$. Electrons do not attach to argon atoms, but only to electronegative impurities. By means of argon purification, detectors with drift lengths greater than 1~m have been obtained \cite{Amerio:2004ze,Bromberg:2013fla}. All these properties are essential to build massive neutrino detectors, at the scale of hundred tons or more. Liquid argon provides a number of other advantages, ensuring not only detector mass, but also a superior neutrino detection capability. The large ionization yield (with an average energy expenditure of about 24 eV to create an electron-ion pair \cite{Miyajima:1974zz}) and the low diffusion of ionization electrons (with longitudinal and transverse diffusion coefficients of about 6 and 16 cm$^2$/ns, respectively, for a 500 V/cm drift field \cite{atrazhev1998transport}) in LAr provide excellent imaging capability and a low energy threshold. Combined with a homogeneous and fully active detection volume, the large ionization yield also ensures accurate calorimetry. Argon de-excitation and electron-ion recombination after the passage of charged particles give rise to an intense scintillation light signal as well. The average energy expenditure to create a scintillation photon in LAr is about 100~eV in the absence of electron-ion recombination, while only 20~eV are needed in the presence of full recombination \cite{Aprile:2009dv}. The prompt scintillation light signal in large LAr detectors has traditionally been used as trigger for neutrino interactions, as veto against cosmic rays, and to provide absolute event timing. In this paper, additional opportunities offered by the scintillation light signal in LAr neutrino detectors are discussed.

After the LAr neutrino detection concept was proposed, the technology was developed by the ICARUS Collaboration over a thirty year long R\&D program. For a historical overview of the LAr technology, see \cite{Marchionni:2013tfa}. The ICARUS T600 detector, with 760 tons, is the largest LAr detector ever built. The ICARUS T600 detector was first operated on the surface and commissioned with cosmic ray data \cite{Amerio:2004ze}. The detector was later exposed to the CNGS neutrino beam at Gran Sasso National Laboratory (Italy) between 2010 and 2012 \cite{Rubbia:2011ft}. Current plans envisage refurbishing ICARUS T600 at CERN \cite{Antonello:2012hf} and possibly moving it to FNAL \cite{Antonello:2013ypa}. European R\&D activities also focus on developing a dual phase argon TPC, with electron amplification occurring in the gaseous volume above the liquid \cite{Badertscher:2013wm}. A large detector of this type is envisaged in the LAGUNA-LBNO proposal to search for the neutrino mass hierarchy and for CP violation in the neutrino sector \cite{Stahl:2012exa,::2013kaa}. A 6$\times$6$\times$6 m$^3$ demonstrator will be built at CERN \cite{DeBonis:1692375}. In the United States, many efforts have started in the last 10 years to resolve technical issues related to building a multi-kton scale LAr detector \cite{Bromberg:2013fla}, as required by the LBNE long-baseline neutrino experiment \cite{Adams:2013qkq}. The ArgoNeuT test-beam project collected neutrino interactions in the Fermilab NuMI beamline in 2009-2010 \cite{Anderson:2011ce}. The MicroBooNE experiment \cite{Chen:2007ae} is a 170 ton LAr TPC in the Fermilab Booster neutrino beamline. The goals of MicroBooNE are to measure low energy neutrino cross sections on argon, to investigate the low energy excess events observed in the MiniBooNE $\nu_{\mu}\to\nu_e$ oscillation search \cite{Aguilar-Arevalo:2013pmq}, and to act as a technology test-bed for future, larger, LAr detectors. MicroBooNE commissioning will start in 2014. Other US R\&D plans on LAr detectors include LArIAT \cite{Adamson:2013/02/28tla}, CAPTAIN \cite{Berns:2013usa} and LAr1-ND \cite{Fleming:2013uaa}. 

In this work, we address how a LAr TPC with enhanced sensitivity to scintillation light may improve detector performance. We focus on two key performance indicators for neutrino oscillation physics, namely neutrino energy reconstruction and neutrino/antineutrino separation. The paper is organized as follows. Section \ref{sec:Simulation} describes the simulation framework used. Results on neutrino energy reconstruction are presented in Sec.~\ref{sec:Energy}. Section \ref{sec:MuonCharge} describes the use of scintillation light to distinguish muon neutrino from muon antineutrino interactions. In Sec.~\ref{sec:Feasibility} we discuss whether large LAr neutrino detectors with sufficiently high efficiency for scintillation light detection appear feasible. We conclude in Sec.~\ref{sec:Conclusions}.

\section{Simulation} \label{sec:Simulation}

In this paper, we use the LArSoft software framework \cite{Church:2013hea} to simulate the charge and light response of a large, ideal, LAr detector. Specifically, we study the LAr response to the interactions of mono-energetic electron neutrinos, muon neutrinos and muon antineutrinos in the 1--6 GeV neutrino energy range. This energy range is inspired by the neutrino flux expectations for the LBNE experiment \cite{Adams:2013qkq}. The detector configuration considered here is ideal because charge drift, light propagation and readout effects are not taken into account.

The GENIE neutrino event generator \cite{Andreopoulos:2009rq} is used by LArSoft to simulate the various neutrino-nucleus interaction processes occurring in this energy regime. The GENIE simulation describes the nuclear effects affecting the initial state of the interaction, the relative rates of the various neutrino-nucleus interaction processes, the final state composition and kinematics at the neutrino interaction vertex, and the final state hadronic interactions occurring within the argon target nucleus. In the following, we consider charged-current (CC) neutrino interactions only, and we classify those according to four categories: neutrino quasi-elastic scattering, neutrino-induced production and subsequent decay of hadronic resonances, neutrino deep inelastic scattering, and neutrino-induced coherent pion production. A minimum of $10^3$ neutrino interactions are generated for each incoming neutrino flavor and energy configuration. In addition to GENIE, we also use a single particle event generator in LArSoft, as discussed in Sec.~\ref{sec:Energy}.

The detector response in LArSoft is modelled according to the Geant4 toolkit for the simulation of the passage of particles through matter \cite{Agostinelli:2002hh,Allison:2006ve}. The detector geometry assumed is that of a LBNE far detector with 10 kton mass. However, our results are independent of detector geometry, provided that the LAr active volume is large enough to fully contain the neutrino interaction products (other than neutrinos). In our simulation, neutrino interactions are generated only in a small (10~cm radius, 10~cm long) cylindrical volume near the center of one of the two 5 kton cryostats envisaged for the LBNE 10 kton detector. As a result, the minimum distance from the neutrino interaction vertex to the active volume boundary is about 3.5~m in the plane transverse to the beam direction, about 20~m in the downstream direction, and about 5~m in the upstream direction. Hadronic interactions (according to the Bertini intranuclear cascade model \cite{Yarba:2012ih}), electromagnetic interactions, and particle decay processes are simulated according to the default Geant4 models included in the {\tt QGSP\_BERT} physics list. The {\tt NeutronTrackingCut} process is disabled in order to follow the tracking of neutrons down to their production threshold. The capture of negatively-charged muons by argon nuclei is also considered. As discussed in Sec.~\ref{sec:MuonCharge}, about a 74\% fraction of stopping $\mu^-$'s are expected to be captured by argon nuclei.

The production of ionization electrons and scintillation photons in LAr is simulated according to NEST (Noble Element Simulation Technique, \cite{Szydagis:2011tk}). In this model, for LAr and for electronic (as opposed to nuclear) energy loss, 17\% of the energy is dissipated into argon excitation, with the remaining 83\% converted into ionization processes, as estimated in \cite{Aprile:2009dv}. Overall, an average energy expenditure of 19.5 eV is assumed to produce one excitation or one ionization reaction in argon \cite{Doke2002}. In the case of energy loss by highly ionizing nuclear fragments, the quenching of ionization and excitation is also accounted for, as some of the energy is dissipated in the form of heat. In a second step, the recombination of electrons is taken into account. This is a function of the so-called linear energy transfer (LET) and of the drift electric field intensity. The former quantity is defined as the quotient between the particle energy deposition per unit path length and the density of the detection medium. In general, the lower the LET and the higher the drift electric field, the lower the recombination rate. The recombination probability $r$ modelled in NEST, for tracks with kinetic energy above \orderof (10 keV), is a modification of Birks' law that was first introduced by Doke \cite{Szydagis:2011tk,Doke:1988dp}:
\begin{equation}
r = \frac{A\frac{dE}{dx}}{1+B\frac{dE}{dx}}+C,\quad C = 1-A/B,
\label{eq:recombination}
\end{equation}
\noindent where $dE/dx$ is the LET expressed in (MeV~cm$^2$/g), and $A$, $B$ and $C$ are empirical parameters that depend on the noble element, density and drift electric field. For LAr, NEST assumes:
\begin{equation}
A = B = 0.07\cdot\mathrm{E_{drift}^{-0.85}},\ C = 0
\label{eq:nestrecombination}
\end{equation}
\noindent where the drift electric field $\mathrm{E_{drift}}$ is expressed in kV/cm. In the following, we take $\mathrm{E_{drift}} = 0.5$ kV/cm and therefore $A = B = 0.126$. Ionization electrons lost into recombination are assumed to produce one scintillation photon each, adding to the scintillation light produced by excitation directly. By simulating both charge and light production along each simulation step and for each Geant4 particle, NEST correctly handles the correlations among the two signals in the LAr detector.

\begin{figure}[t!b!]
\begin{center}
\includegraphics[width=0.60\textwidth]{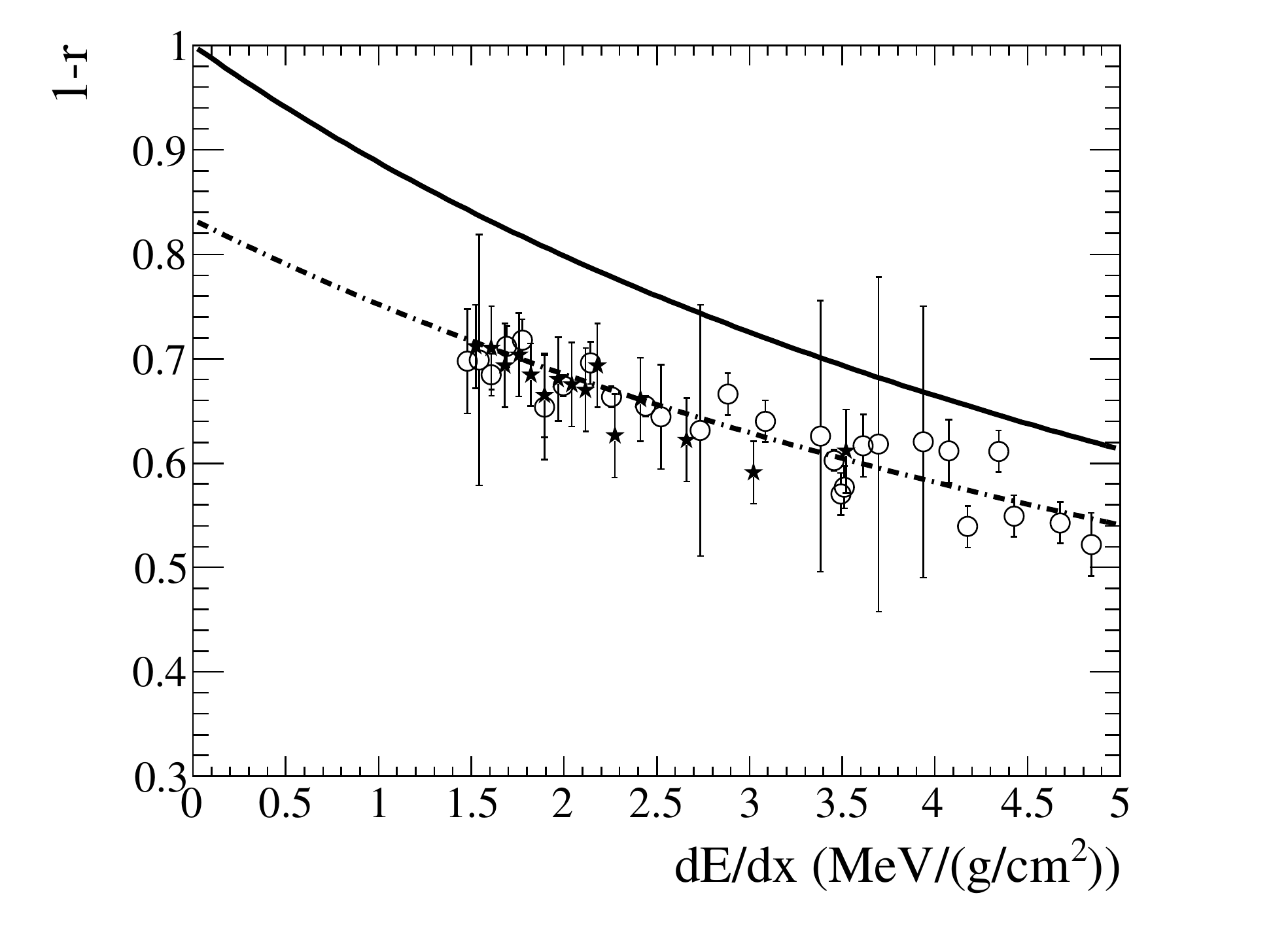} 
\end{center}
\caption{Escape probability $1-r$, where $r$ is the recombination probability, as a function of the linear energy transfer $dE/dx$ in (MeV~cm$^2$/g). The data points are from the ICARUS 3 ton (circles) and ICARUS T600 (stars) LAr TPCs, and are reproduced from \cite{Amoruso:2004dy}. The solid curve is the NEST prediction in Eq.~\protect\ref{eq:nestrecombination}, and the dashed curve is the modified recombination model in Eq.~\protect\ref{eq:modifiedrecombination}.}\label{fig:recombination}
\end{figure}

The results in Secs.~\ref{sec:Energy} and \ref{sec:MuonCharge} are expected to depend somewhat on our simulation assumptions for neutrino-argon interactions, for hadronic interactions in LAr, and for electron-ion recombination. For example, the impact of different hadronic physics models (including the {\tt QGSP\_BERT} Geant4 model adopted here) on hadron shower energy reconstruction in LAr, and therefore on neutrino energy reconstruction in LAr, is discussed in \cite{Stahl:2012exa}. In this work we use the LArSoft framework since it represents the ``state-of-the-art'' for simulating the response of LAr TPCs. The code is publicly available and shared by all US-based LAr neutrino experiments, including ArgoNeuT, MicroBooNE and LBNE. As a result, the LArSoft-based simulation is fast-maturing and has undergone significant validation based on neutrino data from ArgoNeuT already, see \cite{Anderson:2011ce,Acciarri:2013met,Acciarri:2014isz}. In addition, the general-purpose GENIE, Geant4 and NEST simulation tools used by LArSoft have been extensively benchmarked with other data as well. See for example Ref.~\cite{Andreopoulos:2009rq} for GENIE validation results in the few-GeV neutrino energy range of relevance here, although for neutrino interactions on nuclear targets lighter than argon. Concerning the validation of Geant4 hadronic physics models, including the {\tt QGSP\_BERT} model, the reader should consult \cite{Yarba:2012ih}. Finally, a comparison of the NEST recombination model in Eqs.~\ref{eq:recombination} and \ref{eq:nestrecombination} with ICARUS data \cite{Amoruso:2004dy} is shown in Fig.~\ref{fig:recombination}. The ICARUS recombination trend with linear energy transfer is well reproduced by the simulation, but NEST predicts an overall lower recombination probability (that is, a higher escape probability, $1-r$) compared to what suggested by ICARUS data. For this reason, we also extract some results for a modified recombination model that reproduces better ICARUS data, given by the same parametrization in Eq.~\ref{eq:recombination} and the following model parameters:
\begin{equation}
A = 0.05\cdot\mathrm{E_{drift}^{-0.85}},\ B = 0.06\cdot\mathrm{E_{drift}^{-0.85}},\  C = 1/6
\label{eq:modifiedrecombination}
\end{equation}
%

\section{Neutrino energy reconstruction} \label{sec:Energy}

A fundamental event observable that needs to be reconstructed in most neutrino oscillation or neutrino scattering experiments is the energy of the incoming neutrino, \Enu. This can be done reliably only for CC neutrino interactions, so that we restrict the discussion to this type of interaction. Depending on the neutrino energy range, one of two methods is typically used for this purpose, which we call {\it quasi-elastic} and {\it calorimetric} neutrino energy reconstruction, in the following.

\begin{figure}[t!b!]
\begin{center}
\includegraphics[width=0.49\textwidth]{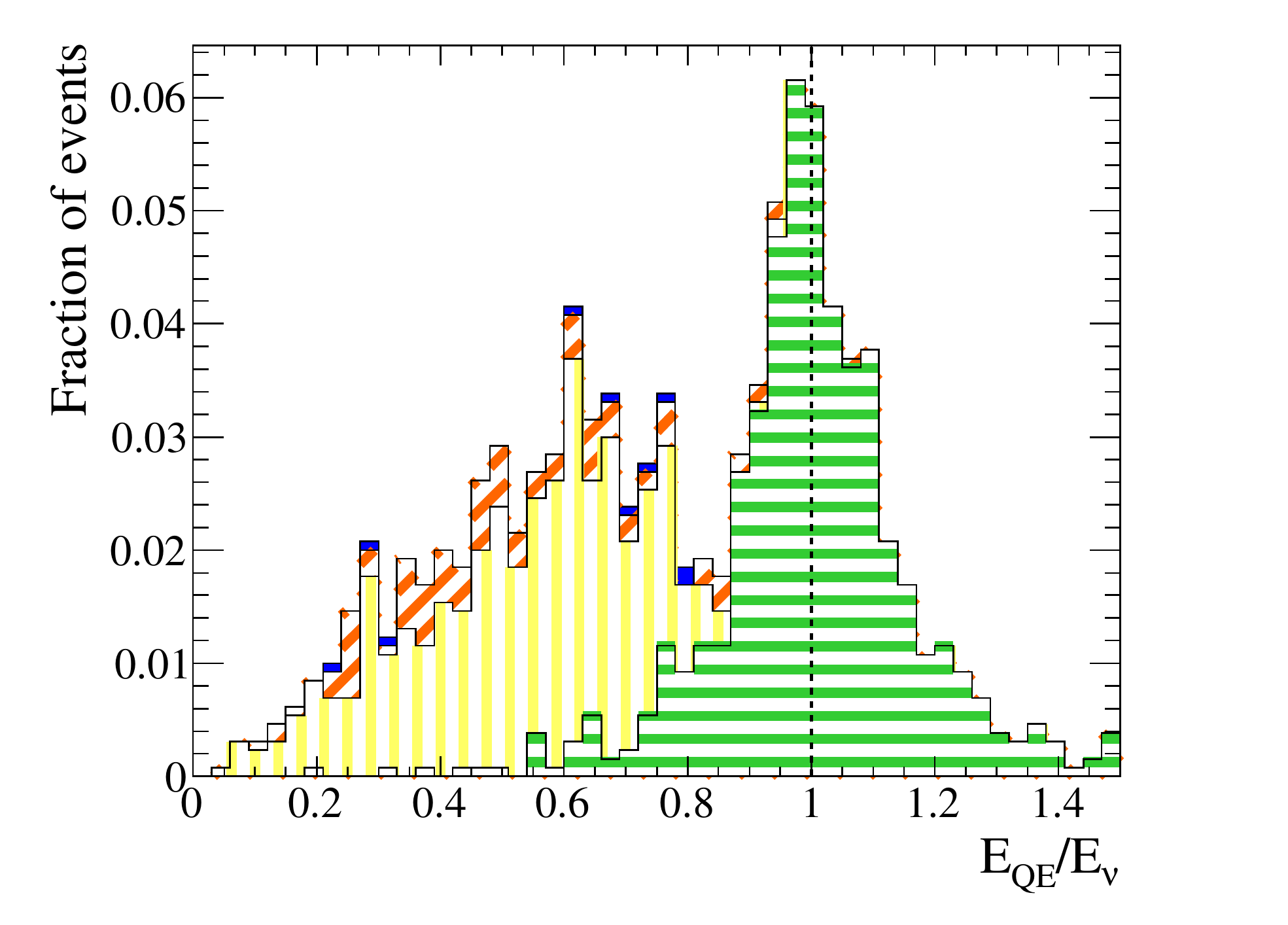} \hfill
\includegraphics[width=0.49\textwidth]{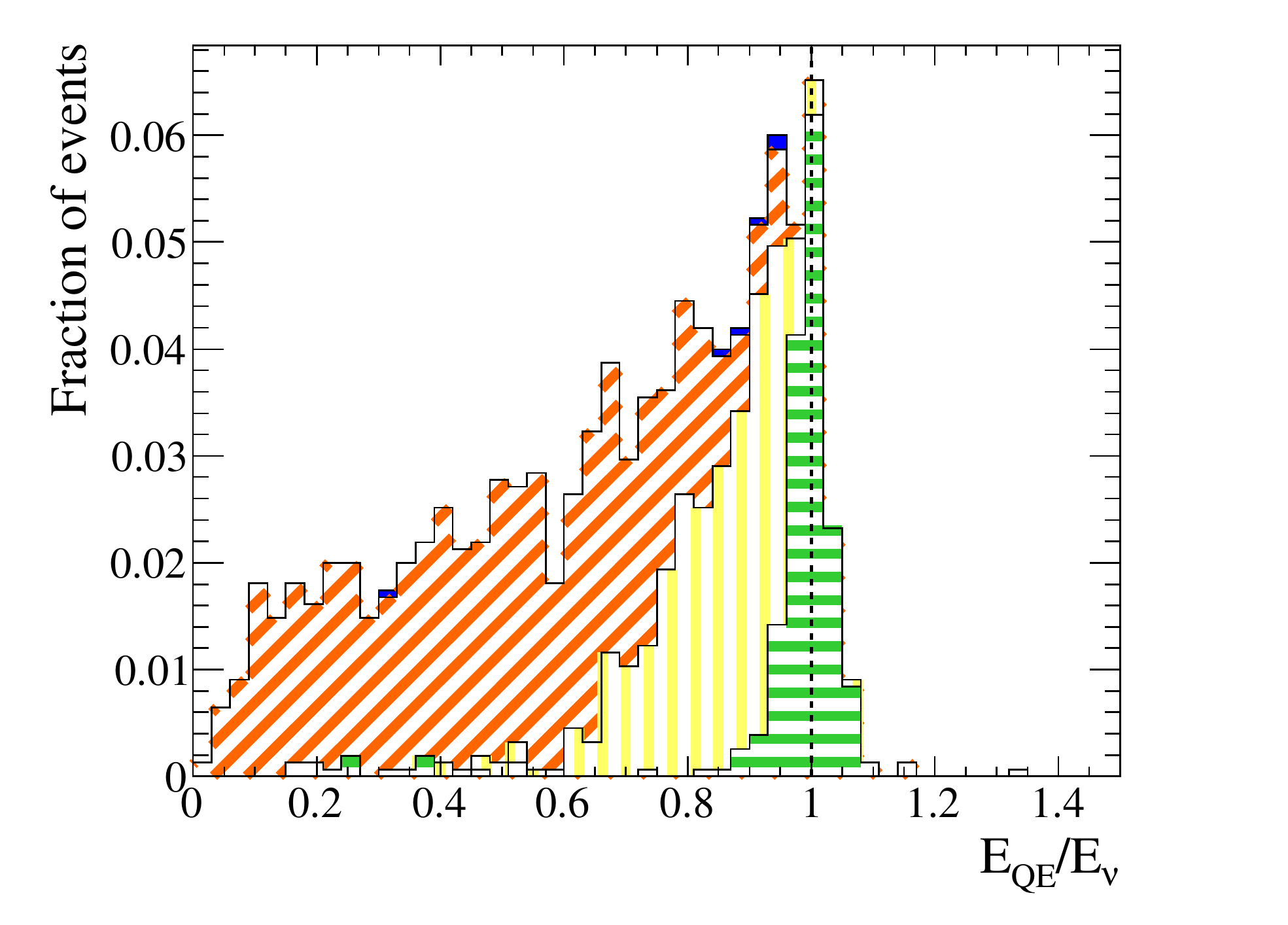} 
\end{center}
\caption{Ratio of reconstructed neutrino energy \Eqe\ using the quasi-elastic interaction assumption (Eq.~\protect\ref{eq:eqe}), divided by the true neutrino energy \Enu, according to simulations. The left (right) panel refers to 1 GeV (4 GeV) $\nu_e$ CC interactions in liquid argon. The green (horizontally hatched), yellow (vertically hatched), orange (diagonally hatched) and blue (filled) histograms indicate quasi-elastic, resonant, deep inelastic and coherent interactions, respectively. }\label{fig:nue_eqe}
\end{figure}

In the quasi-elastic reconstruction, the incoming neutrino energy is estimated by measuring the energy and direction of the outgoing charged lepton. One assumes that the reaction proceeds via the quasi-elastic channel $\nu_l +n\to l^- +p$, that the target neutron is at rest, and that the incoming neutrino direction is known. With these assumptions, the quasi-elastic neutrino energy can be written as: 
\begin{equation}
\Eqe = \frac{1}{2}\frac{m_n^2-(m_p-V)^2-m_l^2+2(m_p-V)E_l}{(m_p-V)-E_l+\lvert\vec{p}_l\rvert\cdot \cos\vartheta_l}
\label{eq:eqe}
\end{equation}
\noindent where $m_n$, $m_p$ and $m_l$ are the neutron, proton and charged lepton mass, respectively, $V$ is the binding energy (about 30 MeV in argon), $E_l$, $\lvert\vec{p}_l\rvert$ and $\vartheta_l$ are the charged lepton total energy, momentum and angle with respect to the neutrino direction, respectively. The ratio of reconstructed neutrino energy \Eqe\ using Eq.~\ref{eq:eqe}, divided by the true neutrino energy \Enu, is shown in Fig.~\ref{fig:nue_eqe} for simulated $\nu_e$ CC interactions in liquid argon. As shown in Fig.~\ref{fig:nue_eqe} and as is well known, Eq.~\ref{eq:eqe} works very poorly for inelastic neutrino interactions such as resonant interactions and deep inelastic scattering. As the energy of the neutrino beam exceeds the few-GeV energy range, it becomes less and less practical to use this method on a CC inclusive sample, given that the fraction of CC interactions that are quasi-elastic decreases rapidly. Alternatively, one could use Eq.~\ref{eq:eqe} only on a sub-sample enriched in quasi-elastic neutrino interactions, for neutrino detectors providing quasi-elastic/inelastic interaction separation capability and at the cost of increased analysis complexity. However, the use of Eq.~\ref{eq:eqe} is not optimal also in this case, given that one would be left with a much reduced statistical sample. Also, from Fig.~\ref{fig:nue_eqe}, one can see that even for quasi-elastic events and for perfect reconstruction of the charged lepton kinematics (as assumed in the figure), the incoming neutrino energy cannot be perfectly measured. The reason is that the target nucleons are bound in argon nuclei, and rather than being at rest as assumed in Eq.~\ref{eq:eqe}, they move with typical Fermi momenta of order 250 MeV/c.

\begin{figure}[t!b!]
\begin{center}
\includegraphics[width=0.49\textwidth]{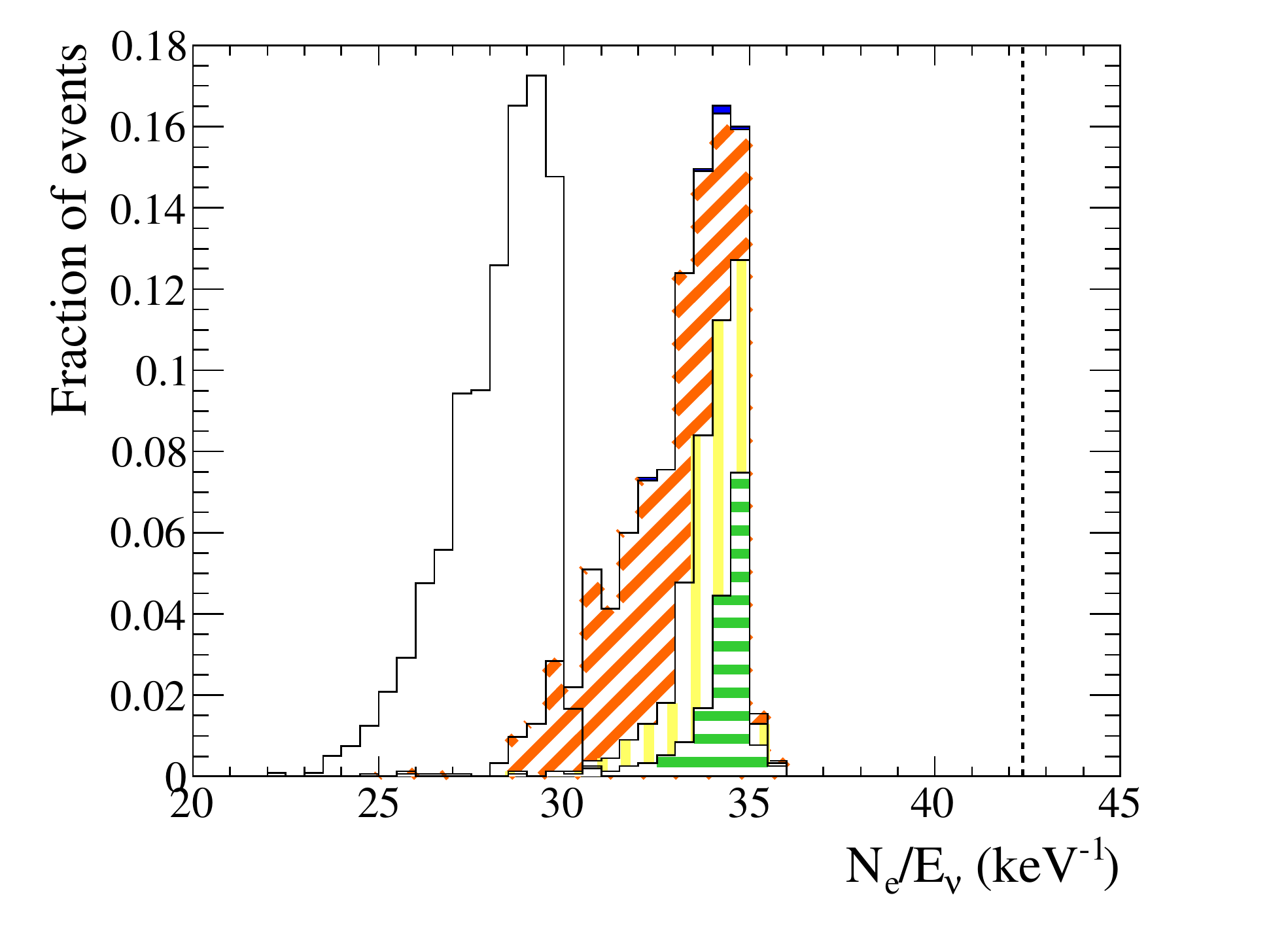} \hfill
\includegraphics[width=0.49\textwidth]{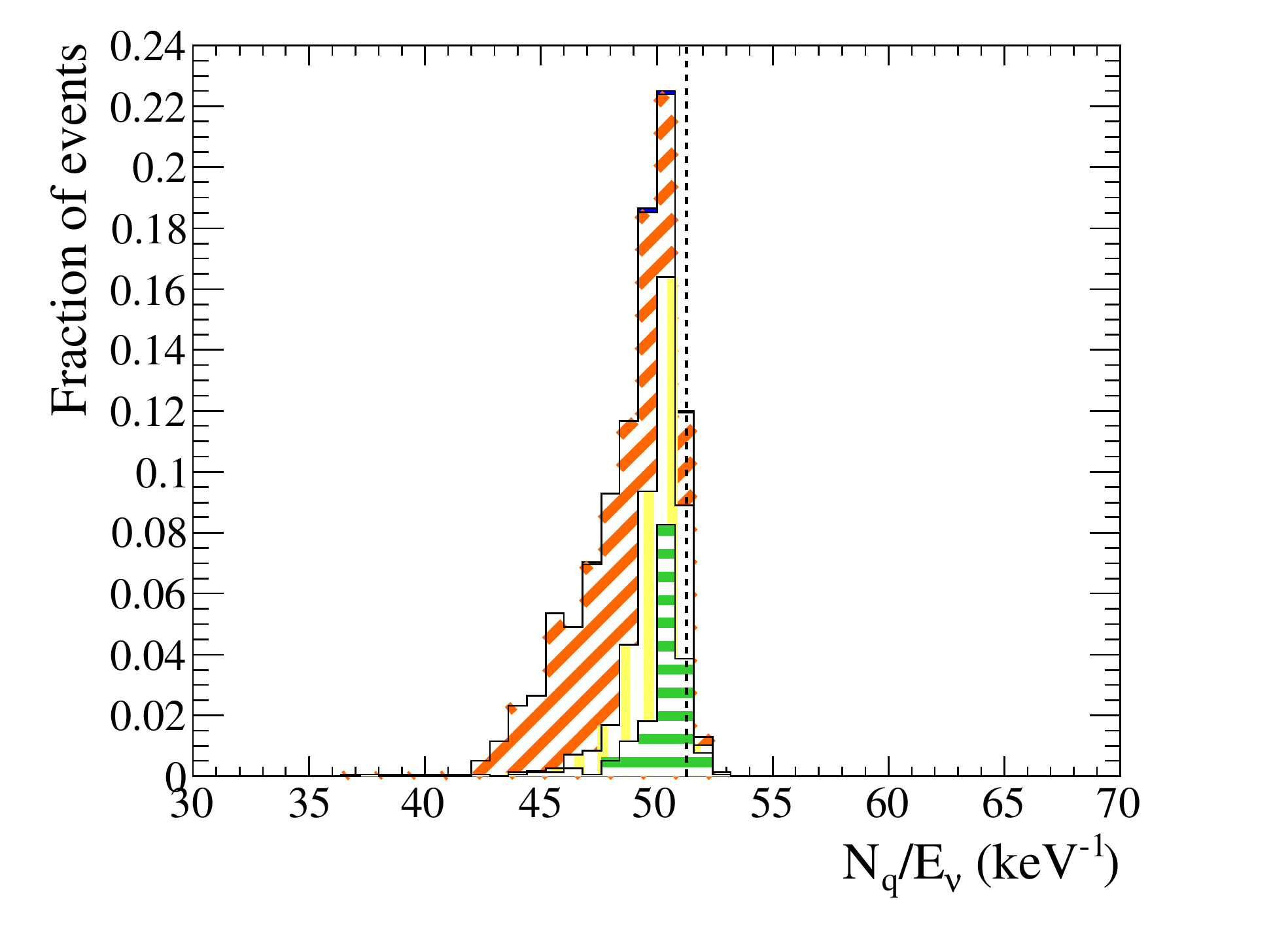} 
\includegraphics[width=0.49\textwidth]{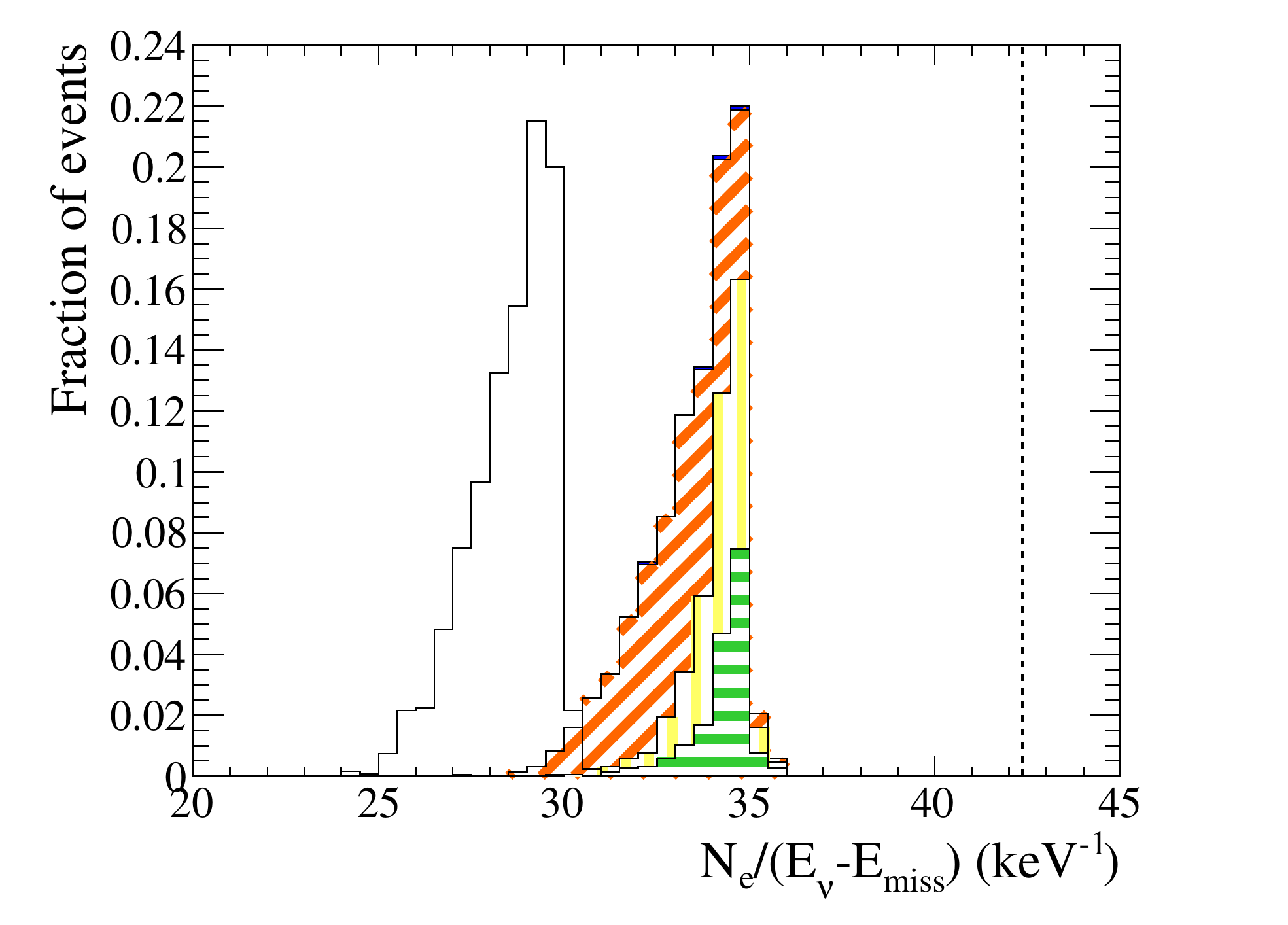} \hfill
\includegraphics[width=0.49\textwidth]{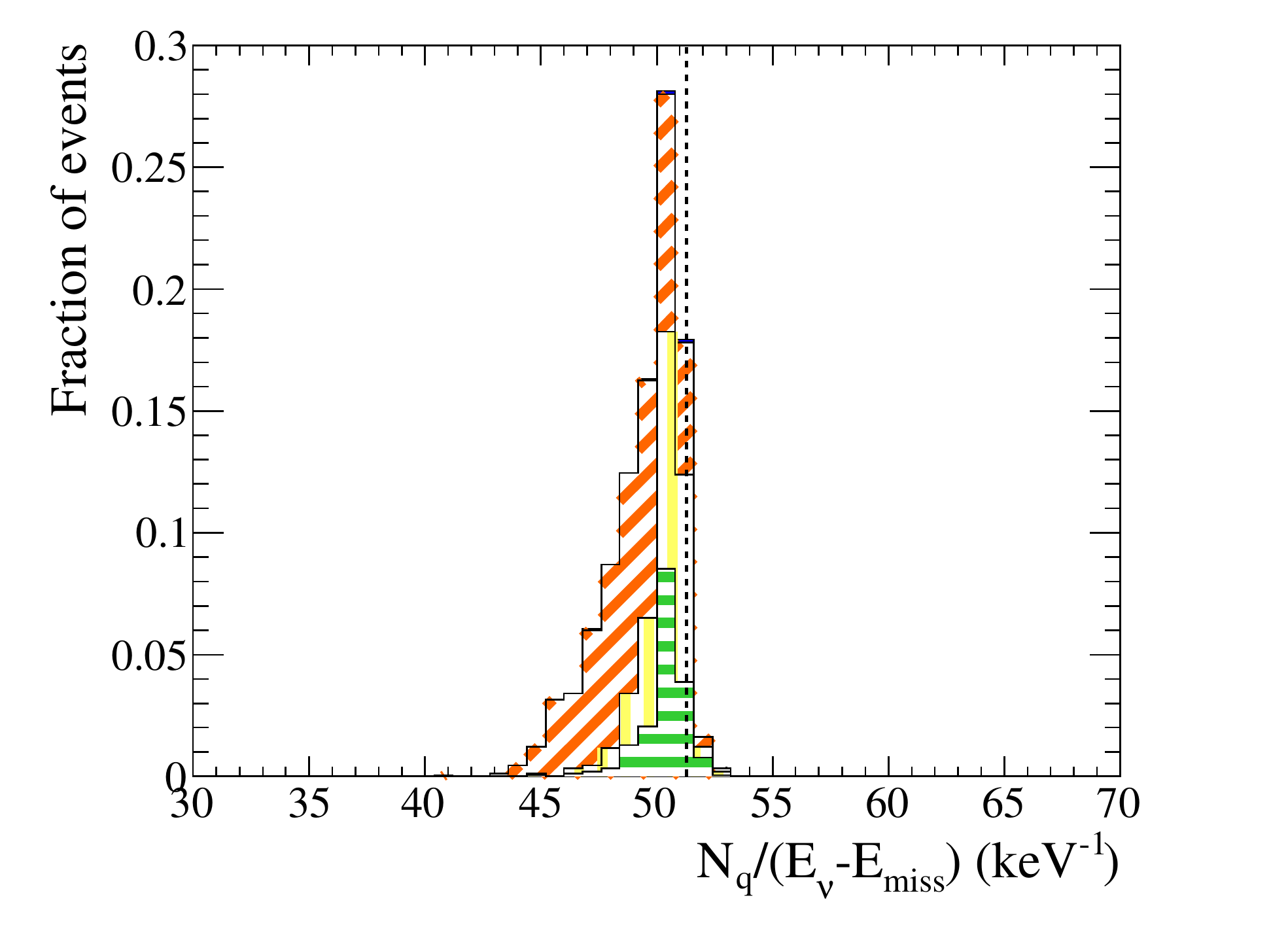} 
\end{center}
\caption{Calorimetric neutrino energy reconstruction for simulated 4 GeV $\nu_e$ CC interactions, for various observables defined in the text. The histogram color and hatching convention is the same as in Fig.~\protect\ref{fig:nue_eqe}. Top left panel: ratio of the number of ionization electrons \Ne\ divided by true neutrino energy \Enu; top right: ratio of the number of quanta (scintillation photons or ionization electrons) \Nq\ divided by \Enu\ (see Sec.~\protect\ref{subsec:Energy_Recombination}); bottom left: ratio of \Ne\ divided by (\Enu-\Emiss), the true energy of the primary neutrino minus the energy escaping the detector in the form of secondary neutrinos (see Sec.~\protect\ref{subsec:Energy_SecondaryNus}); bottom right: ratio of \Nq\ divided by (\Enu-\Emiss), see Sec.~\protect\ref{subsec:Energy_Resolution}. The vertical line at 42.4 keV$^{-1}$ in the top left and bottom left panels shows the ionization yield expected in the absence of electron-ion recombination. The line at 51.3 keV$^{-1}$ in the top right and bottom right panels is the expected photon plus electron yield. All histograms are obtained with the NEST recombination model in Eq.~\protect\ref{eq:nestrecombination}, except for the two additional empty histograms on the left panels, obtained with the modified recombination model of Eq.~\protect\ref{eq:modifiedrecombination}.}\label{fig:nue_ecalo}
\end{figure}

In the calorimetric neutrino energy reconstruction, one simply adds all the energy deposited in the fully active detector volume. As the basis for the visible energy measurement in a LAr detector, the total charge produced by ionization and reaching the readout planes is used. In practice, several effects limit the precision of this calorimetric measurement, namely \cite{Stahl:2012exa}:  
\begin{enumerate}
\item nuclear effects in neutrino interactions, such as those due to the nuclear potential or to the reinteractions of the interaction products;
\item non-deposited energy that is carried away by secondary neutrinos;
\item particle (other than secondary neutrino) leakage out of the active volume;
\item quenching of the ionization (and excitation) produced by heavy nuclear fragments and other highly ionizing particles;
\item electron-ion recombination;
\item electron attachment to electronegative impurities along drift;
\item electronic noise.
\end{enumerate}
The resolution of the calorimetric measurement is limited by the fluctuations in the above effects. Some of these effects have been studied in previous simulation works, see for example  \cite{Stahl:2012exa,Para:2002rm,Harris:2003si}. In the idealized simulation discussed here, we purposefully do not consider the contributions to the resolution from particle leakage out of the active volume, from electron attachment, and from electronic noise (effects 3, 6, 7). Previous studies by the ICARUS Collaboration have demonstrated that both attachment and noise contributions to the energy resolution can be made small. On the one hand, a Monte-Carlo sample of low-energy (E~$<$~50~MeV) electrons is used in \cite{Amoruso:2003sw} to estimate an electronic noise contribution to the energy resolution of order 0.3\%$/\sqrt{\rm E~(GeV)}$. On the other hand, the energy-independent contribution to the energy resolution from electron attachment is estimated to be at most 1.2\% in ICARUS, using 0.05--5~GeV photons and for electron lifetimes of order 1~ms \cite{Ankowski:2008aa}. The simulated detector active volume surrounding the neutrino interaction is large enough to avoid particle leakage, as it is expected to be the case for a large fraction of the detector volume in future, very massive, LAr detectors. Neglecting leakage, attachment and noise effects, the expected calorimetric performance of 4 GeV electron neutrino interactions in LAr can be seen in the top left panel of Fig.~\ref{fig:nue_ecalo}, where \Ne\ indicates the total number of ionization electrons produced in the LAr active volume and escaping recombination. As can be seen in the figure, the expected energy resolution is markedly better than the typical resolutions obtained with the quasi-elastic method in Fig.~\ref{fig:nue_eqe}, and less dependent on the neutrino interaction details.

In the rest of this section, we aim to study the effect on calorimetric neutrino energy reconstruction coming from the remaining effects listed above, namely: nuclear effects, quenching of charge and light signals associated to nuclear fragments, energy carried away by secondary neutrinos, and electron-ion recombination (effects 1, 2, 4, 5). We will argue that the latter two effects (4 and 5) can, at least partially, be corrected for in a LAr detector with enhanced sensitivity to scintillation light (Secs.~\ref{subsec:Energy_Recombination} and \ref{subsec:Energy_SecondaryNus}). The neutrino energy resolution of an ideal LAr detector affected only by nuclear and charge/light quenching effects will be estimated in Sec.~\ref{subsec:Energy_Resolution}.


\subsection{Impact of electron-ion recombination} \label{subsec:Energy_Recombination}

\begin{figure}[t!b!]
\begin{center}
\includegraphics[width=0.49\textwidth]{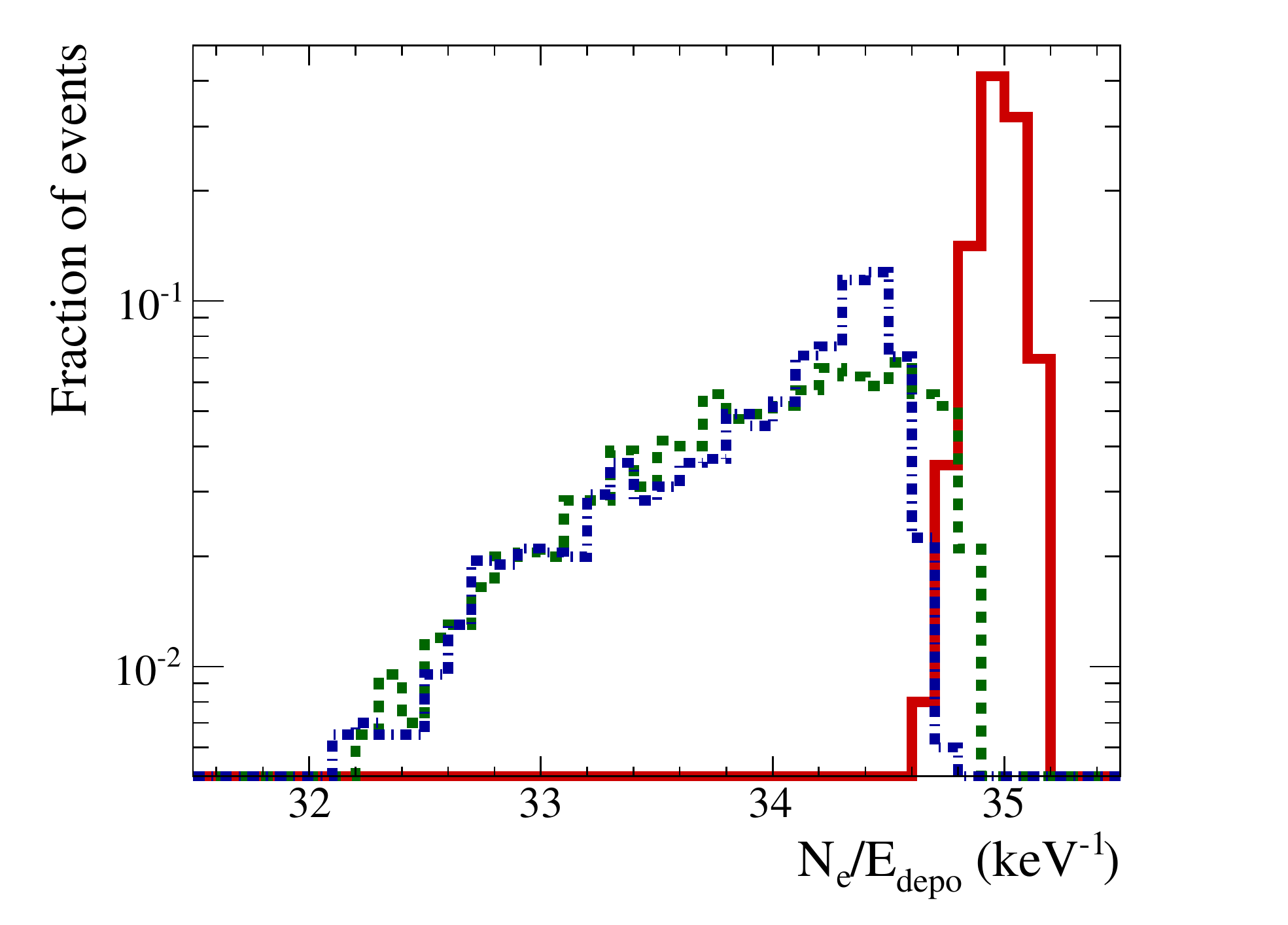} \hfill
\includegraphics[width=0.49\textwidth]{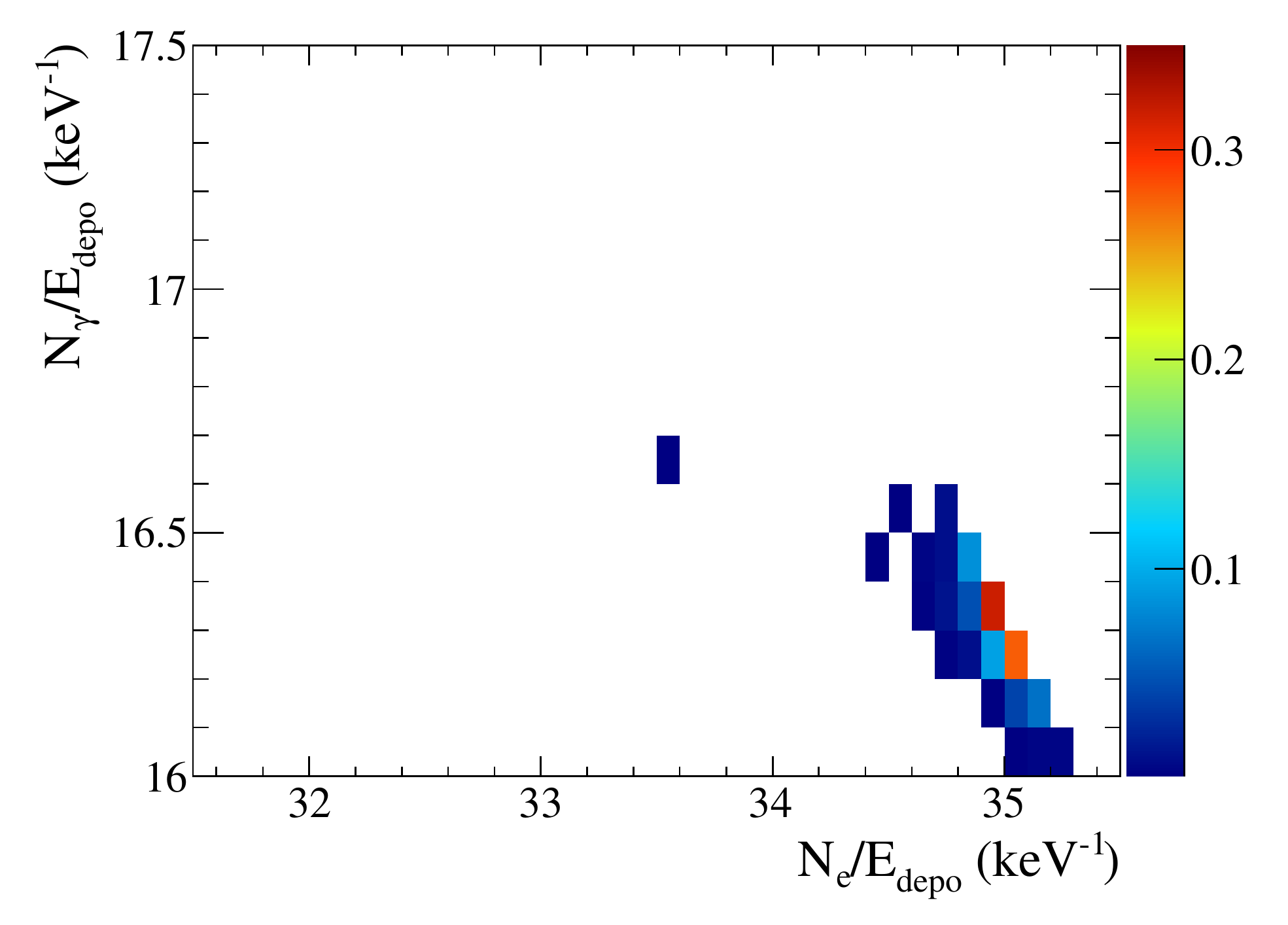} \hfill
\includegraphics[width=0.49\textwidth]{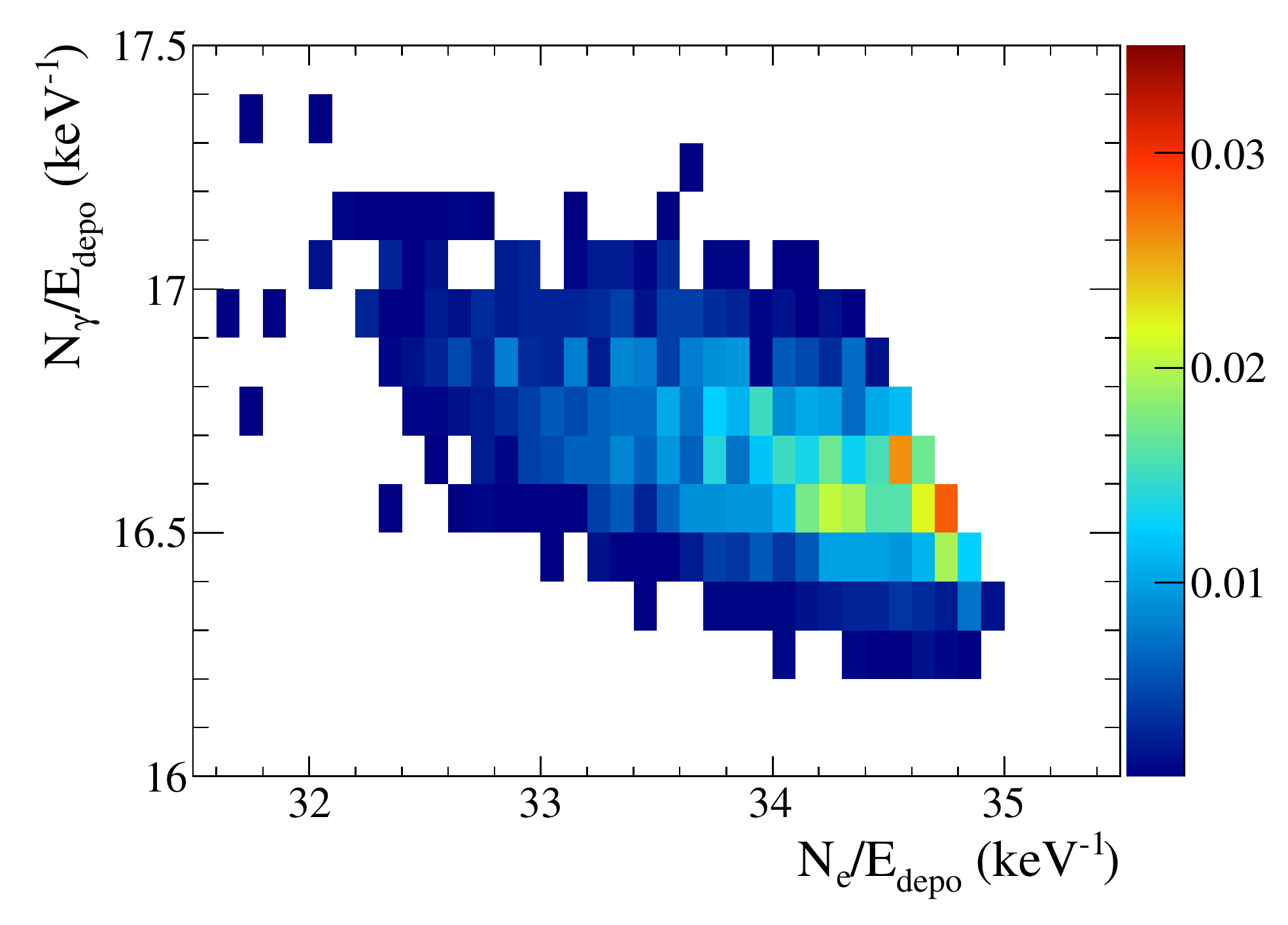} \hfill
\includegraphics[width=0.49\textwidth]{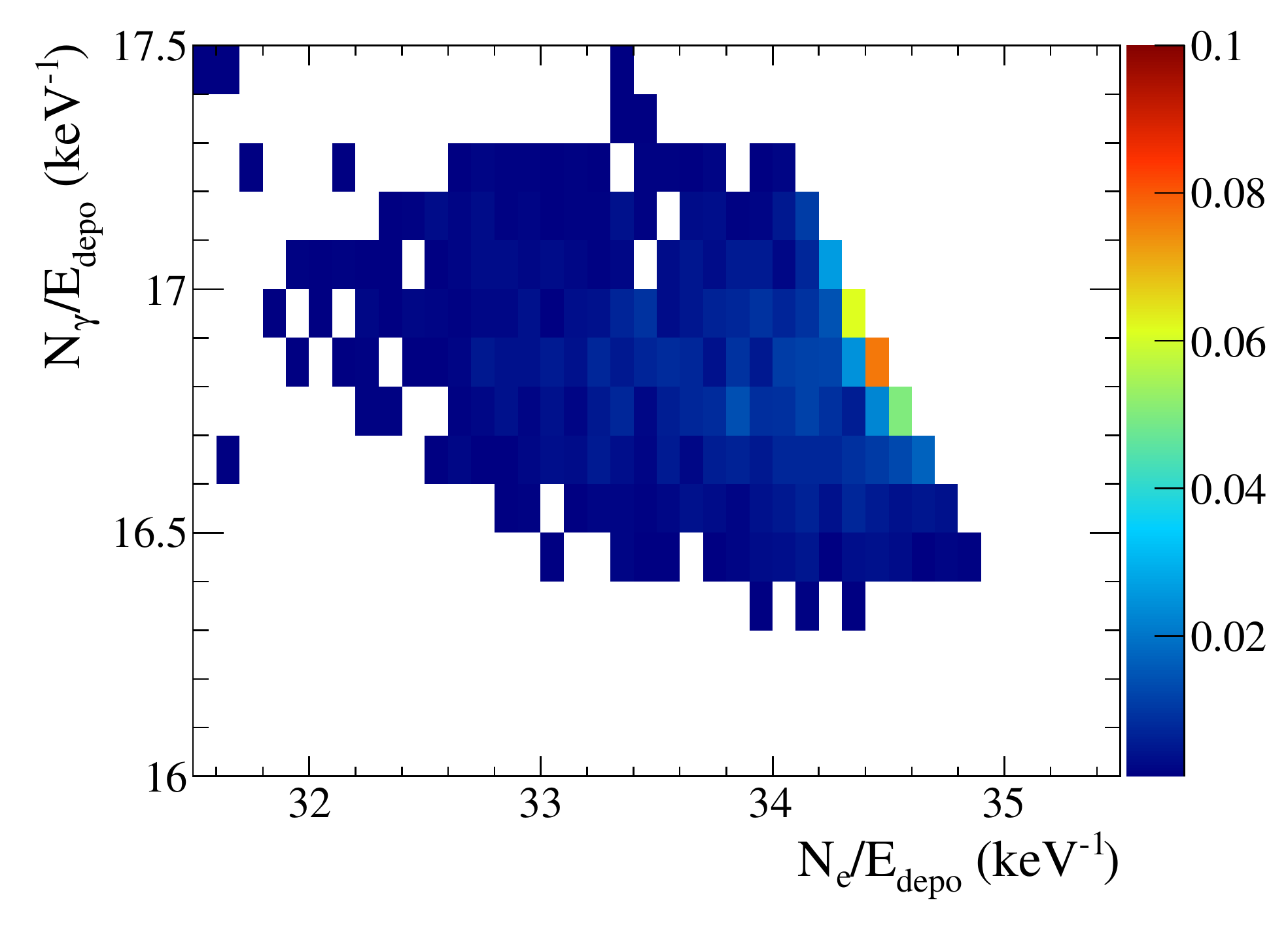} 
\end{center}
\caption{Top left panel: ratio of the number of ionization electrons \Ne\ divided by deposited energy \Edepo. The ratio is shown for 500 MeV kinetic energy primary particles generated in liquid argon: electrons (red solid line), positively-charged pions (green dashed line), protons (blue dash-dotted line). The other three panels in the figure show the number of scintillation photons \Ngamma\ produced per unit deposited energy \Edepo\ versus \Ne/\Edepo, separately for 500 MeV kinetic energy electrons (top right), $\pi^+$'s (bottom left), protons (bottom right).}\label{fig:singleparticle_nelec_nphot}
\end{figure}

The top left panel of Fig.~\ref{fig:singleparticle_nelec_nphot} shows the ratio of \Ne\ to deposited energy (\Edepo) for 500 MeV kinetic energy single particles of different types. As mentioned above, we do not simulate electron attachment nor electronic noise effects. In addition, by normalizing to \Edepo, we explicitly do not account for non-deposited energy stemming from nuclear effects, secondary neutrinos or particle leakage out of the detector. In other words, Fig.~\ref{fig:singleparticle_nelec_nphot} is only affected by quenching and by recombination effects.  

Figure~\ref{fig:singleparticle_nelec_nphot} shows that the average \Ne/\Edepo\ values for electrons, positively-charged pions and protons are different. This is largely due to the different average recombination probabilities for the three particle types. As shown in Eq.~\ref{eq:recombination}, the recombination probability depends upon the particle linear energy transfer. For example, a proton track has a linear energy transfer (or energy deposition per unit path length) that is higher, on average, than the one of an electron of the same kinetic energy. As a consequence, the proton track suffers from a higher average recombination probability, compared to the electron. Given that neutrino interactions can produce a variety of particle types over a range of energies, the charge signal in a LAr calorimeter will respond somewhat differently to different interactions of neutrinos of the same energy. In addition, we can see (Fig.\ref{fig:singleparticle_nelec_nphot}, top left) that the charge distribution has some spread associated to it even in the case of monochromatic particles of the same particle type. This has to do with fluctuations in the recombination process. Because of the stochastic nature of recombination, two particles of the same type and depositing the same amount of energy in the LAr volume will, in general, produce a different number of ionization electrons escaping recombination. Overall, from the differences and spreads of the single-particle charge distributions in the top left panel of Fig.\ref{fig:singleparticle_nelec_nphot}, we expect the recombination contribution to the calorimetric neutrino energy resolution to be at the few percent level.

In principle, the charge and light signals can be combined to reduce recombination effects, ensuring a more compensating LAr calorimeter response. The realization that an average increase in scintillation light (in that case, as the electric field was lowered) could compensate a corresponding average decrease in charge was first proposed for liquid argon and liquid xenon in 1978 \cite{Kubota:1978oha}. Charge and light signals have been combined on an event-by-event basis to reduce the recombination fluctuations in xenon detectors for double beta decay searches \cite{Conti:2003av,Alvarez:2012hu} and for gamma-ray astronomy \cite{Aprile:2007qd}, improving the energy resolution.

The expected anti-correlation between charge (\Ne) and light (\Ngamma) signals due to recombination in LAr for 500 MeV single electrons, positively-charged pions and protons is separately shown for each particle type in Fig.~\ref{fig:singleparticle_nelec_nphot}. It is clear from Fig.~\ref{fig:singleparticle_nelec_nphot} that a better estimator for the total deposited energy is provided by \Nq$\equiv$\Ne+\Ngamma, the total number of quanta (either ionization electrons or scintillation photons) produced. We also find that the anti-correlation between charge and light is almost perfect for electrons, while this is not the case for protons and especially $\pi^+$'s. This smearing is due to charge and light quenching effects, suppressing the \Ne\ and \Ngamma\ signals associated to nuclear projectiles, which are in turn produced by the hadronic interactions of pions and protons.

The top right panel in Fig.~\ref{fig:nue_ecalo} shows the ratio of \Nq\ to the neutrino energy \Enu, for 4 GeV electron neutrino interactions. The suppression of the recombination effects improves the calorimeter response, as can be seen by comparing with the top left panel in the same figure. The vertical lines at 42.4 and 51.3 keV$^{-1}$ in the top left and top right panels of Fig.~\ref{fig:nue_ecalo} show the signals expected for \Enu = \Edepo\ (which is only approximately true) and for no recombination. About a 22\% recombination probability for 4 GeV electron neutrino interactions in a LAr volume filled with a 500 V/cm drift field can be extracted from the top left panel, in agreement with the NEST assumption in Fig.~\ref{fig:recombination}. On the other hand, the charge plus light signal shown in the top right panel is consistent with no recombination, as expected. 

In practice, the recombination correction would not be as straightforward as described above. On one hand, Poisson fluctuations in the detected scintillation light signal are increased by the low light detection efficiency in a LAr detector (see Sec.\ref{sec:Feasibility} for overall efficiency considerations). On the other hand, position-dependent effects in the detected light signal per unit deposited energy, unavoidably present, need to be corrected for at the percent level or better, for this method to be applicable. Position-dependent effects can come from the geometry of the light collection system, from the light absorption in LAr ({\it e.g.,} by nitrogen, see Sec.~\ref{subsec:Energy_SecondaryNus}) and from additional light losses caused by materials in front of the light collection system ({\it e.g.,} charge readout wires). Michel electrons from cosmic ray muons stopping in LAr can provide a useful calibration sample to understand this correction \cite{Kaleko:2013eda}. For LAr detectors operated underground, which are exposed to a highly suppressed cosmic ray flux compared to surface detectors, other calibration strategies may have to be considered. 

We have also examined to what extent our results are sensitive to the details of the recombination model assumed in the simulation, in light of the low recombination rate modeled in NEST (see Fig.~\ref{fig:recombination}). We focus on energy observables that do not attempt to correct for recombination, since these are the most sensitive to recombination assumptions. The empty histogram in the top left panel of Fig.~\ref{fig:nue_ecalo} shows the \Ne/\Enu\ distribution for 4 GeV $\nu_e$ CC interactions of all types obtained with the modified recombination model of Eq.~\ref{eq:modifiedrecombination}. As expected, the net collected charge is lower in this case, corresponding to an average recombination probability of about 33\%. However, the spread in the distribution is unaffected by the change in the recombination model, yielding a compatible calorimetric neutrino energy resolution in the two cases. The same conclusion holds for all neutrino energies and for all energy observables discussed in this work.


\subsection{Impact of missing energy in the form of secondary neutrinos} \label{subsec:Energy_SecondaryNus}

Another factor limiting the calorimetric neutrino energy reconstruction of any detector is the fluctuation introduced by the undetected energy per event that is carried away by secondary neutrinos. Secondary neutrinos typically arise from the decays of muons and charged pions in the neutrino interaction final state. Charged pions are produced in inelastic neutrino interactions. Muons are produced either directly in CC interactions of muon neutrinos or antineutrinos, or indirectly as pion decay products.

\begin{figure}[t!b!]
\begin{center}
\includegraphics[width=0.60\textwidth]{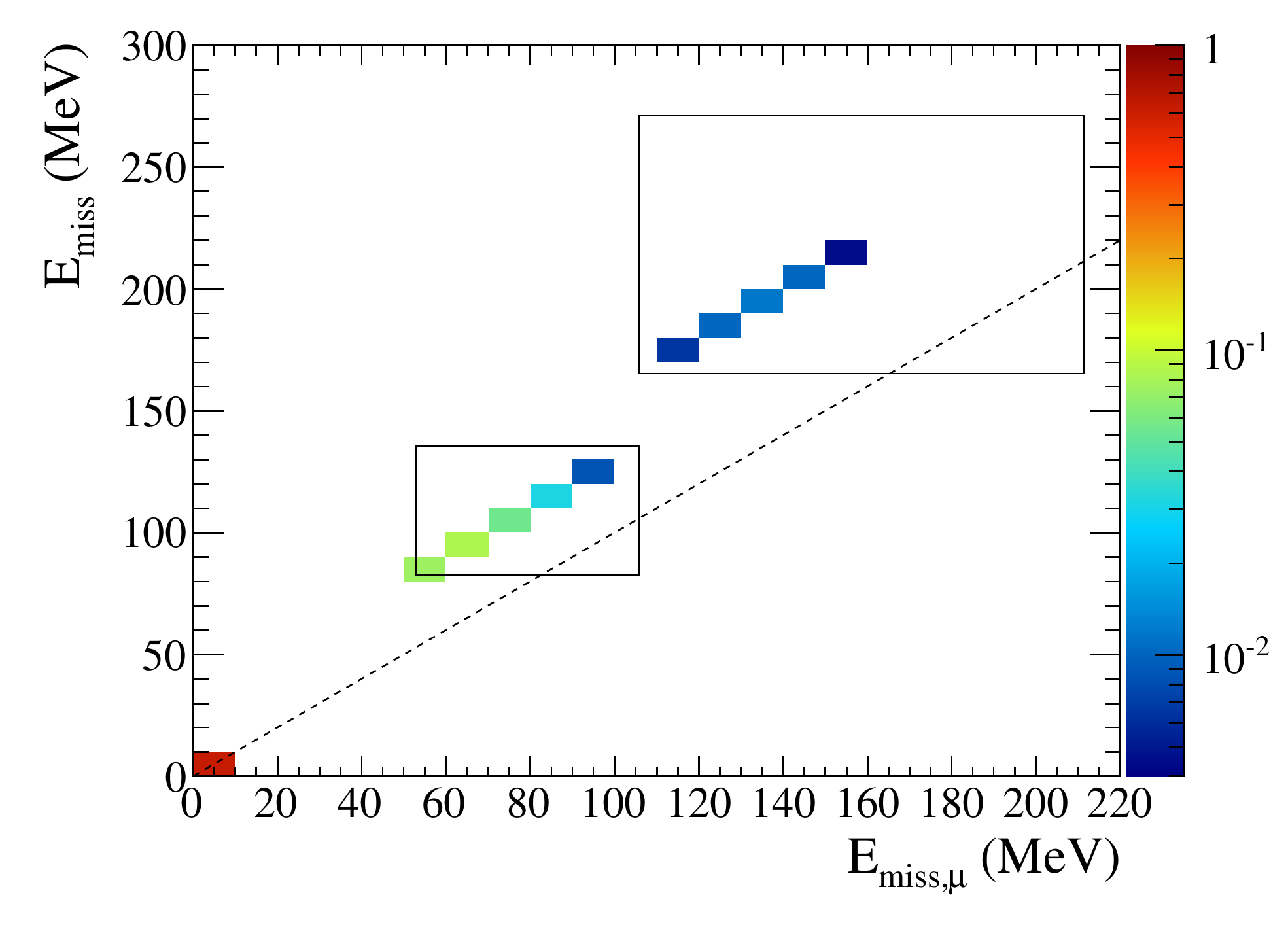} 
\end{center}
\caption{Total missing energy in the form of secondary neutrinos, \Emiss, versus the missing energy due to secondary neutrinos produced in muon decays only, \Emissmu. The plot refers to 4 GeV $\nu_e$ CC interactions. Three populations can be seen for increasing \Emiss\ values: interactions with 0 muon decays (at \Emiss = 0), 1 muon decay (left box) and 2 muon decays (right box). In order to highlight that \Emiss$>$\Emissmu\ in the presence of muon decays, the \Emiss$=$\Emissmu\ diagonal is also drawn with a dashed line.}\label{fig:emiss}
\end{figure}

We show in Fig.~\ref{fig:emiss} that the knowledge of the missing event energy in the form of secondary neutrinos produced in muon decays (\Emissmu) provides a very good estimate of the total energy carried away by secondary neutrinos in the event (\Emiss), for electron neutrino interactions. In this case, muons are mostly produced via pion decay, and the dominant process producing secondary neutrinos is:

\begin{equation}
\pi^{+}\to\mu^{+}+\nu_{\mu},\quad \mu^{+}\to e^{+}+\overline{\nu}_{\mu}+\nu_e
\label{eq:secondaryneutrinos}
\end{equation}

We obtain that the average $\mu^+$ multiplicity per event in 4 GeV $\nu_e$ interactions is about 0.42. On the other hand, very few $\pi^-$'s decay to produce negatively-charged muons, with an average $\mu^-$ multiplicity per event of less than 0.01 for the same type of events. As shown in Fig.~\ref{fig:emiss}, for events with no no muons, \Emiss\ is zero, and no correction is needed. For events with a muon multiplicity of one (two), \Emiss\ is about 30 (60) MeV higher than \Emissmu. This can be readily understood from Eq.~\ref{eq:secondaryneutrinos}, accounting for the energy carried away by the $\nu_{\mu}$'s produced in $\pi^+$ decays at rest. As a result, if a detector is capable to tag Michel electrons from $\mu^+$ decay at rest and to measure their energy, one can in principle infer \Emiss, the total energy carried away by secondary neutrinos, on a event-by-event basis. 

The ICARUS Collaboration has already demonstrated the capability of a large LAr detector to detect Michel electrons, see \cite{Amoruso:2003sw}. In that case, the ionization signal is used to reconstruct the muon track, the Michel electron track near the muon end-point (about 10 cm long in LAr), and the Michel electron energy in a simple event topology: cosmic ray muons stopping in the detector. In the following, we describe an alternative, possibly simpler and more efficient, way to reconstruct Michel electrons based on scintillation light timing.  

\begin{figure}[t!b!]
\begin{center}
\includegraphics[width=0.60\textwidth]{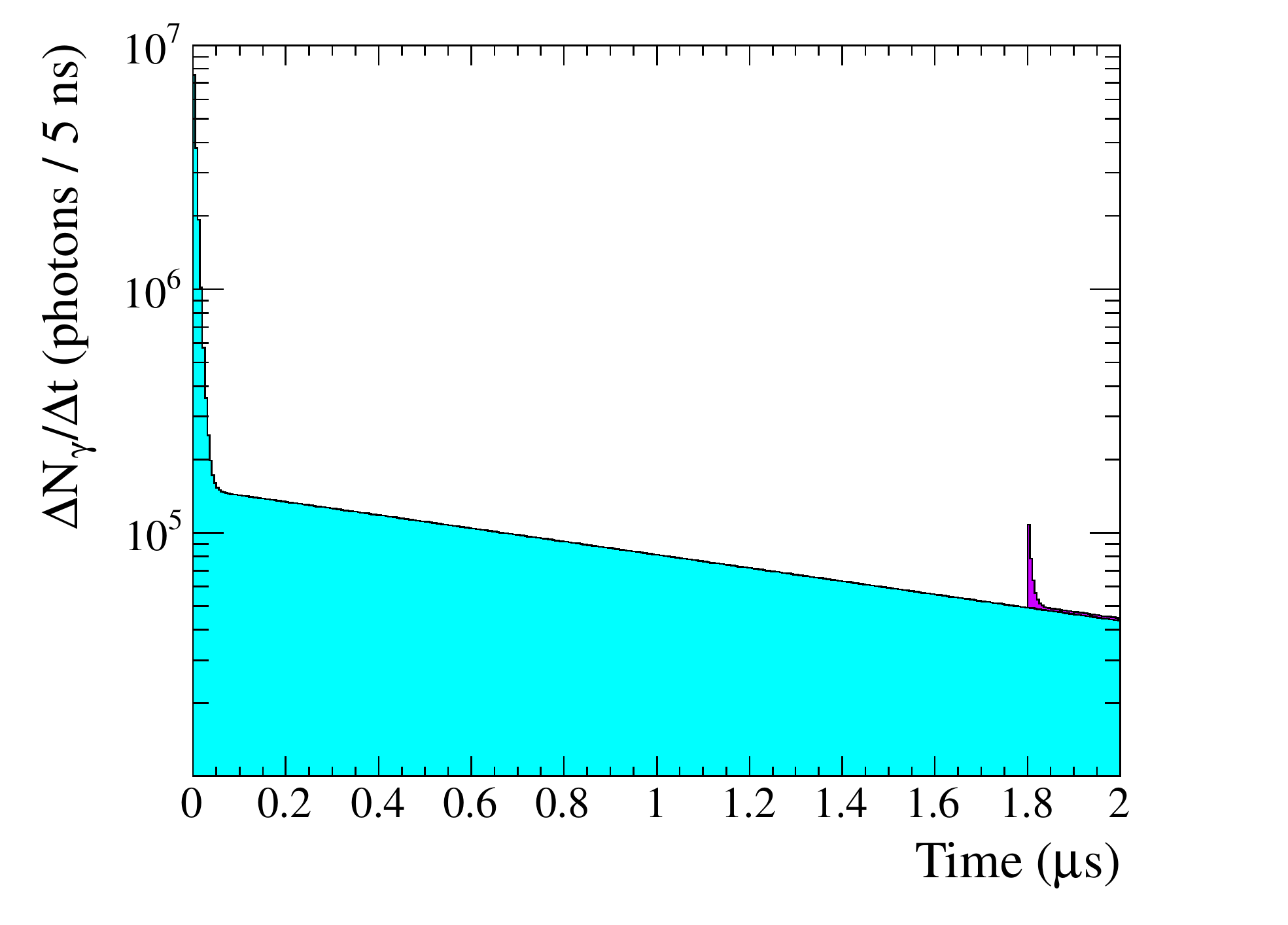} 
\end{center}
\caption{Number of scintillation photons per unit time $\Delta$\Ngamma/$\Delta\mathrm{t}$ as a function of time expected for a 4 GeV neutrino interaction. The cyan component corresponds to the light produced by the primary neutrino interaction, while the violet portion corresponds to the light associated to a 30 MeV decay electron occurring 1.8~$\mu$s after the interaction.}\label{fig:ScintTimeDependence}
\end{figure}

The scintillation light in a LAr detector provides a highly complementary event signature with respect to the ionization signal. While no imaging can be easily performed from scintillation, its prompt nature makes it very appealing to detect delayed coincidences. Figure \ref{fig:ScintTimeDependence} shows the expected number of scintillation photons produced as a function of time, for a 4 GeV neutrino interaction followed by a 30 MeV Michel electron. As shown in Fig.~\ref{fig:singleparticle_nelec_nphot}, about 16 scintillation photons per keV are produced in LAr for a 500 V/cm field and for the NEST recombination model in Eq.~\ref{eq:nestrecombination}, yielding about half a million photons from a 30 MeV electron. Even more light is produced for the modified recombination model of Eq.~\ref{eq:modifiedrecombination}, with about 21 scintillation photons per keV. While the produced signal is very intense (see Sec.~\ref{sec:Feasibility} for light detection efficiency considerations), a complication arises from the non-trivial time dependence of the scintillation light in LAr. Scintillation light arises from the emission of excited diatomic molecules (excimers) of argon. The time evolution of scintillation light from LAr is well described by a two-component model \cite{Hitachi:1983zz}. The fast component, accounting for about 23\% of the total light for minimum ionizing particles, is described by a decay constant of $(7\pm 1)$ ns and is attributed to the transitions from the low excited singlet molecular state $^1\Sigma_u^+$ to the ground state. The slow (late) component, accounting for the remaining 77\%, is characterized by a decay constant of $(1.6\pm 0.1)$ $\mu$s and is due to triplet ($^3\Sigma_u^+$) transitions to ground state \footnote{The 0.23:0.77 ratio of singlet to triplet states was measured at zero drift field and may vary with applied electric field in LAr, see \cite{PhysRevB.20.3486}.}. Given the similarity between the slow decay time constant and the muon decay lifetime, and as shown in Fig.~\ref{fig:ScintTimeDependence}, the slow scintillation light produced by the primary neutrino interaction overlaps in time with the fast light produced by the delayed Michel electron, complicating the detection of the latter. Despite this complication, Michel electrons can in principle be efficiently tagged and their energy accurately reconstructed, for sufficiently high light detection efficiencies. In addition, the late light fraction may not be as high as 0.77 in practice. First, the contamination by impurities in LAr is known to suppress the late light, with little effect on the prompt yield. The WArP Collaboration has measured that a nitrogen contamination of a few parts per million (ppm), that is at the level expected for current-generation LAr TPCs, is sufficient to considerably suppress the late light at production \cite{Acciarri:2008kv}. On the other hand, such a level of nitrogen impurity concentration is not expected to produce a significant amount of light absorption \cite{Jones:2013bca}. Second, the intensity ratio of fast to slow light has been found to increase over the 0.23:0.77 ratio discussed above, for increasing deposited energy densities \cite{Hitachi:1983zz}. 

The expected improvement in calorimetric neutrino energy reconstruction, after correcting for missing energy in the form of secondary neutrinos and for 4 GeV $\nu_e$ interactions in LAr, is shown in Fig.~\ref{fig:nue_ecalo}. The bottom left panel in the figure shows \Ne/(\Enu-\Emiss), characterized by a narrower distribution compared to the \Ne/\Enu\ ratio shown in the top left panel.


\subsection{Achievable neutrino energy resolution} \label{subsec:Energy_Resolution}

\begin{figure}[t!b!]
\begin{center}
\includegraphics[width=0.65\textwidth]{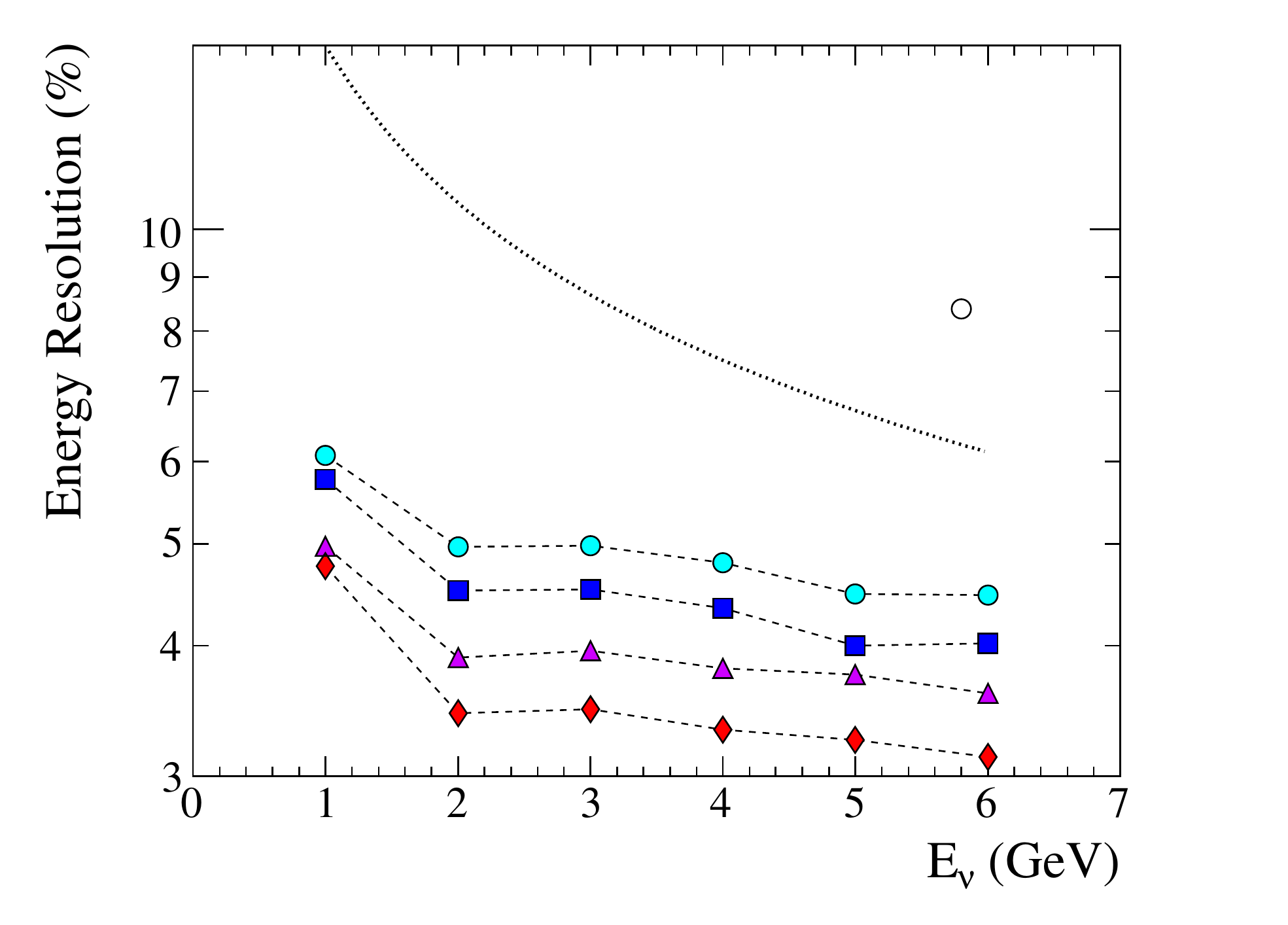}
\end{center}
\caption{Neutrino energy resolution as a function of true neutrino energy \Enu, using the various energy observables defined in the text and shown in Fig.~\protect\ref{fig:nue_ecalo}. The cyan circles assume using \Ne\ alone, as in the top left panel of Fig.~\protect\ref{fig:nue_ecalo}; the blue squares refer to \Nq, that is correcting for electron-ion recombination, as in the top right panel; the violet triangles are obtained by correcting for missing energy in the form of secondary neutrinos, that is using \Ne/(\Enu-\Emiss) as in the bottom left panel; the red diamonds include both the recombination and the secondary neutrinos corrections, as for \Nq/(\Enu-\Emiss) in the bottom right panel. The neutrino energy resolution assumed in LBNE \cite{Adams:2013qkq} and LBNO \cite{Stahl:2012exa} physics studies are also shown, with a dotted curve and an empty circle, respectively.}\label{fig:nue_energyresolution}
\end{figure}

A summary of the resolution expected for the various calorimetric neutrino energy reconstruction methods discussed in the text are shown in Fig.~\ref{fig:nue_energyresolution}, for electron neutrino interactions in LAr in the 1--6 GeV energy range. The fractional resolution given is simply defined as the RMS divided by the mean for the distributions shown in Fig.~\ref{fig:nue_ecalo}. The fractional resolution improves mildly with increasing energy for all observables discussed, by a factor of about 1.4 as the neutrino energy is increased from 1 to 6 GeV. Assuming that resolution effects due to energy leakage out of the detector, electron attachment along drift, and electronic noise can all be made negligible, a resolution of 4.8\% RMS for 4 GeV $\nu_e$ interactions appears achievable. As discussed, corrections can be made to further improve this resolution. Correcting for the missing energy in the form of secondary neutrinos (Sec.~\ref{subsec:Energy_SecondaryNus}) appears to be somewhat more important than ensuring a more compensating calorimeter response via the recombination correction (Sec.~\ref{subsec:Energy_Recombination}). However, both corrections yield non-negligible improvements. If both corrections are implemented, an ideal resolution as low as 3.3\% RMS is achievable for 4 GeV $\nu_e$ interactions, see the bottom right panel in Fig.~\ref{fig:nue_ecalo} and Fig.~\ref{fig:nue_energyresolution}. The latter is an essentially irreducible value, due to nuclear effects and to charge and light quenching for highly ionizing tracks.

These results can be compared with the $\nu_e$ energy resolution of 15\%/$\sqrt{\mbox{E (GeV)}}$ assumed for LBNE physics studies \cite{Adams:2013qkq}. The energy resolution assumed by LBNE is dominated by fluctuations in the hadronic showers energy measurement, taken to be 30\%/$\sqrt{\mbox{E (GeV)}}$ from ICARUS estimates \cite{Rubbia:2011ft} based on FLUKA simulations \cite{Ferrari:2000wu}, and considering that final state hadrons carry about 40\% of the energy of the incoming neutrino in the 1--6 GeV neutrino energy range. A similar, and slightly worse, resolution is assumed in LBNO physics studies, with an overall 8.4\% RMS resolution for $\nu_e$ CC interactions in the 1--10 GeV range and at an average energy of 5.8 GeV, based on Geant4 simulations with the {\tt QGSP\_BERT} physics list \cite{Stahl:2012exa}. Our simulation studies show that these assumptions may in principle be improved by up to a factor of 2--3, particularly at low (1--2 GeV) neutrino energy.

It is worth noticing that our simulation provides better calorimetric neutrino energy resolution than both ICARUS and LBNO simulations, even assuming charge-only information (cyan circles in Fig.~\ref{fig:nue_energyresolution}) and particularly at low energies. This difference cannot be attributed to the effects that are not simulated here, namely particle leakage out of the detector, electron attachment, or electronic noise (see Sec.~\ref{sec:Simulation}). As discussed in Sec.~\ref{subsec:Energy_Recombination}, we have also verified that the relatively low recombination rate modelled in NEST (Eq.~\ref{eq:nestrecombination}) does not affect the energy resolution, compared to a simulation that better reproduces ICARUS data (Eq.~\ref{eq:modifiedrecombination}). The likely cause of this discrepancy has to do with different hadronic physics assumptions, and particularly with the treatment of low-energy ($<$20 MeV) neutrons. Neutrons contribute to the calorimetric measurement in two ways \cite{Ferrari:2000wu}: through kinetic energy transferred to charged particle recoils, and particularly through gammas produced in $(n,n^{\prime})$ and $n$ capture reactions. In simulations, neutrons are often not tracked once they fall below some pre-defined kinetic energy threshold or above some propagation time, in order to reduce CPU time. For example, the default {\tt QGSP\_BERT} model in Geant4 neglects neutrons beyond $>10~\mu$s times via the so-called {\tt NeutronTrackingCut} process, despite that neutron thermalization times in LAr can extend up to $\mathcal{O}$(ms). This approximation is not entirely correct, since LAr TPCs should integrate signals up to timescales equivalent to their full TPC drift time, also of order $\mathcal{O}$(ms). For this reason, the {\tt NeutronTrackingCut} has been disabled in our simulation, see Sec.~\ref{sec:Simulation}. We have verified that maintaining the {\tt NeutronTrackingCut} would have produced a 12\% resolution for 1 GeV $\nu_e$ CC interactions, to be compared with 6\% in Fig.~\ref{fig:nue_energyresolution}. In addition to neutron modelling, other hadronic physics aspects may affect neutrino energy resolution results from simulations, and comparisons among them.

\section{Muon charge identification} \label{sec:MuonCharge}

Another key performance indicator for a detector studying neutrino oscillations is the capability to distinguish neutrino from antineutrino interactions. This is particularly true for proposed experiments to search for CP violation in the neutrino sector, where neutrino and antineutrino oscillation probabilities are typically to be compared. The most effective way to distinguish neutrino from antineutrino CC interactions relies on measuring the curvature in a magnetic field of the final state lepton. This has been accomplished both in large magnetized iron calorimeters \cite{Michael:2008bc} as well as in relatively small tracking detectors \cite{Altegoer:1997gv, Abe:2011ks}. Magnetized solutions for totally active detectors, such as LAr TPCs, have also been studied within the context of proposals for future neutrino facilities, see \cite{Antonello:2013ypa,Abe:2007bi,Adey:2013pio}. This solution may be, however, technically too challenging or prohibitively expensive. Phenomenological studies on non-magnetized solutions to perform neutrino/antineutrino separation have been discussed in \cite{Huber:2008yx}, and include the muon capture technique discussed in the following.


\subsection{Muon capture} \label{subsec:MuonCharge_DecayCapture}

For a muon neutrino beam, information on the neutrino/antineutrino content can be obtained by exploiting $\mu^-$ capture on nuclei in a non-magnetized detector. This technique has already been applied to a carbon-based detector, see \cite{AguilarArevalo:2013hm}. The advantage of argon over lighter nuclei is the much higher capture rate.

In particular, the rate for a stopping $\mu^-$ to either decay or be captured by a argon nucleus is given by:
\begin{equation}
\frac{1}{\tau} = \frac{1}{\tau_d}+\frac{1}{\tau_c}
\label{eq:muminusprocess}
\end{equation}

\noindent where $\tau_d = 2.197$ $\mu$s is the muon lifetime in vacuum. For argon, the Geant4 simulation assumes a muon capture rate of $1/\tau_c = 1.30$ $\mu$s$^{-1}$, resulting in a muon lifetime in argon of $\tau = 0.570$ $\mu$s and in a muon capture fraction $f_c = \tau/\tau_c = 0.74$. This expected value is essentially compatible with the value $\tau = (0.537\pm 0.032)$ $\mu$s measured in \cite{Suzuki:1987jf}. 

\begin{figure}[t!b!]
\begin{center}
\includegraphics[width=0.49\textwidth]{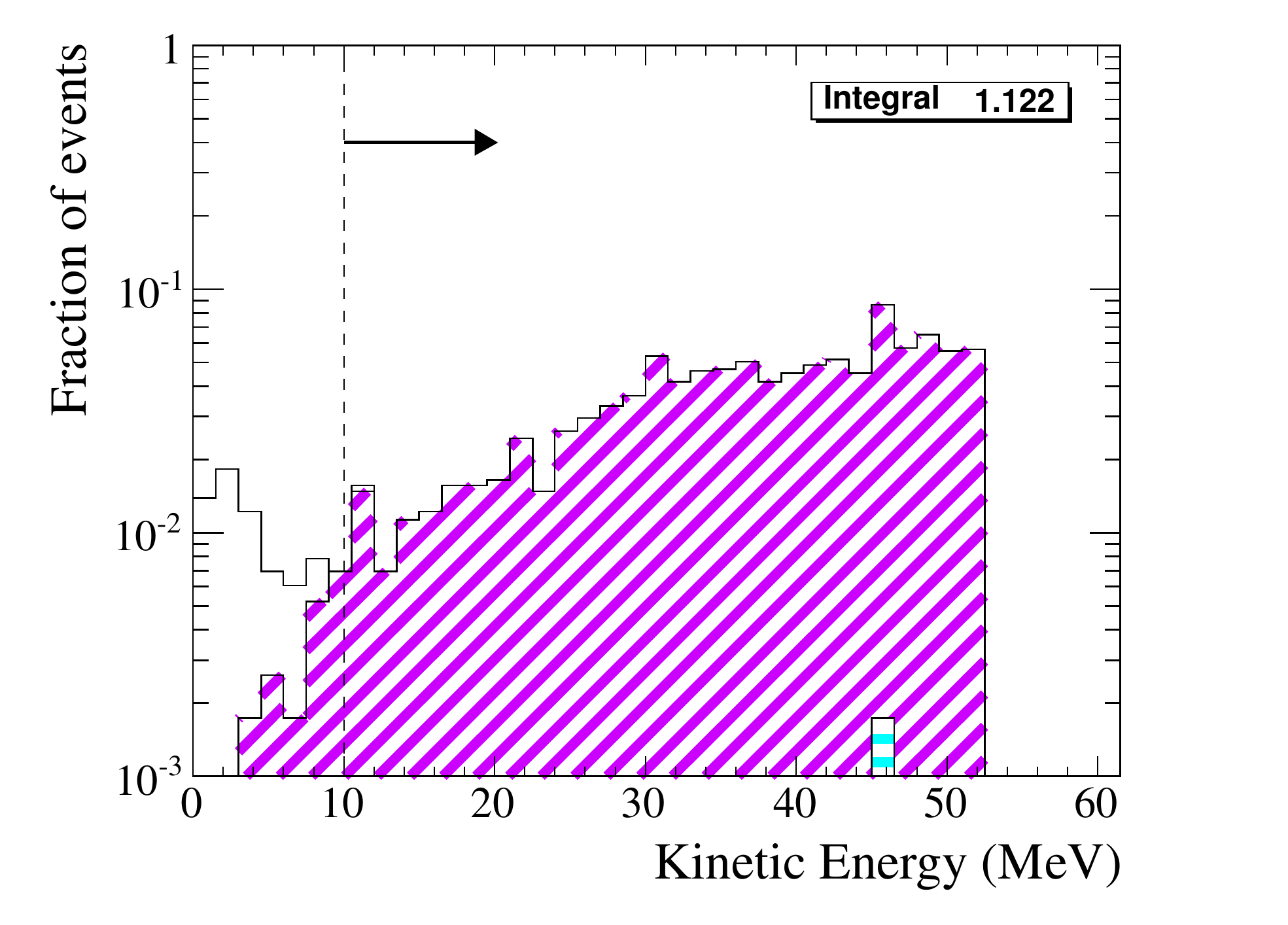} \hfill
\includegraphics[width=0.49\textwidth]{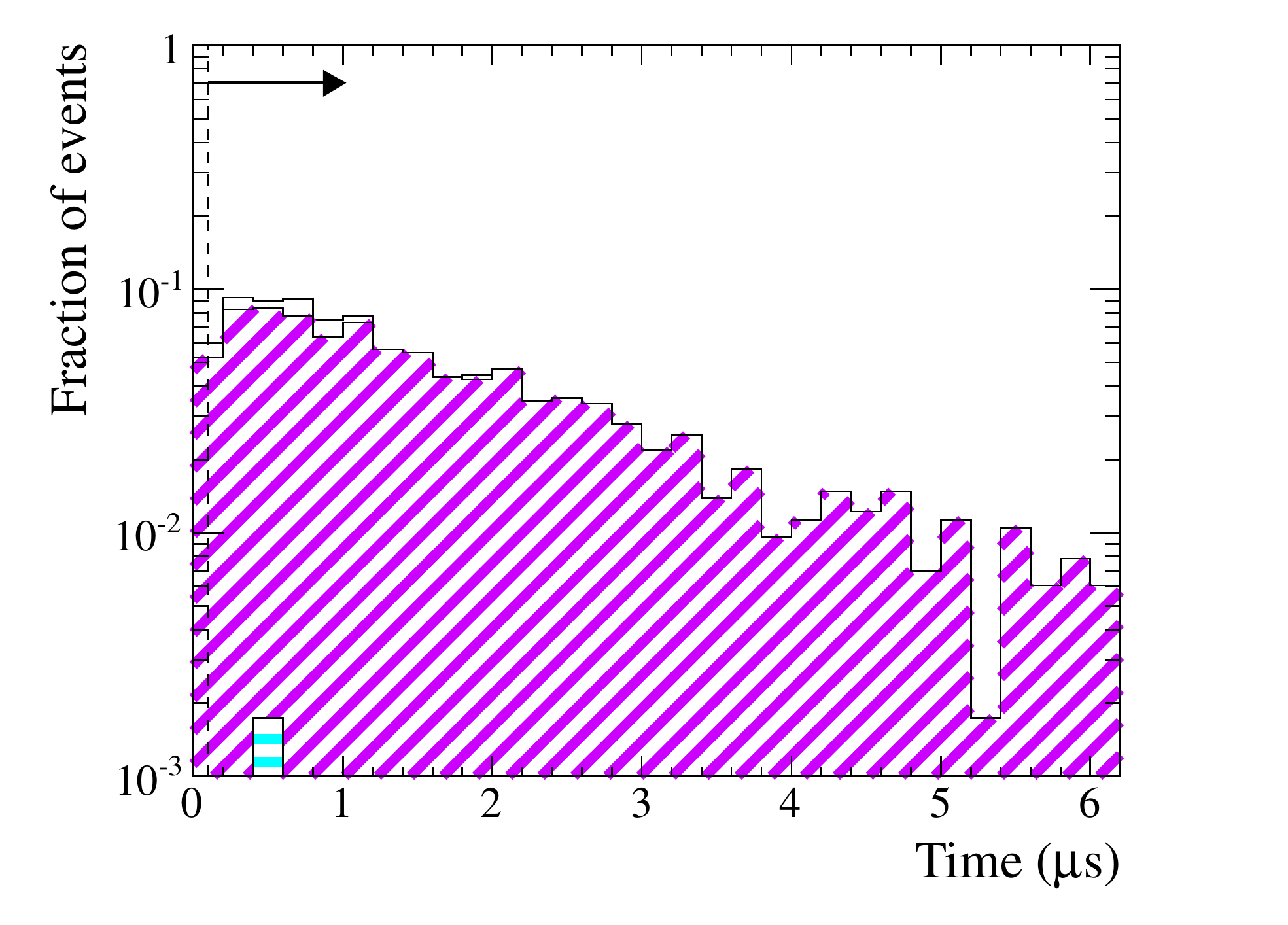} 
\end{center}
\caption{Kinetic energy (left panel) and initial time (right) for particles produced in muon decay or capture processes, originated from the interactions of 4 GeV muon antineutrinos in liquid argon. The cyan (horizontally hatched) component shows electrons from $\mu^-$ decay, the violet (diagonally hatched) component shows positrons from $\mu^+$ decay, while other particles are shown in white. The vertical lines and arrows indicate the energy and timing requirements to be fulfilled by Michel electron candidates.}\label{fig:numubar_decayenergytime}
\end{figure}

\begin{figure}[t!b!]
\begin{center}
\includegraphics[width=0.49\textwidth]{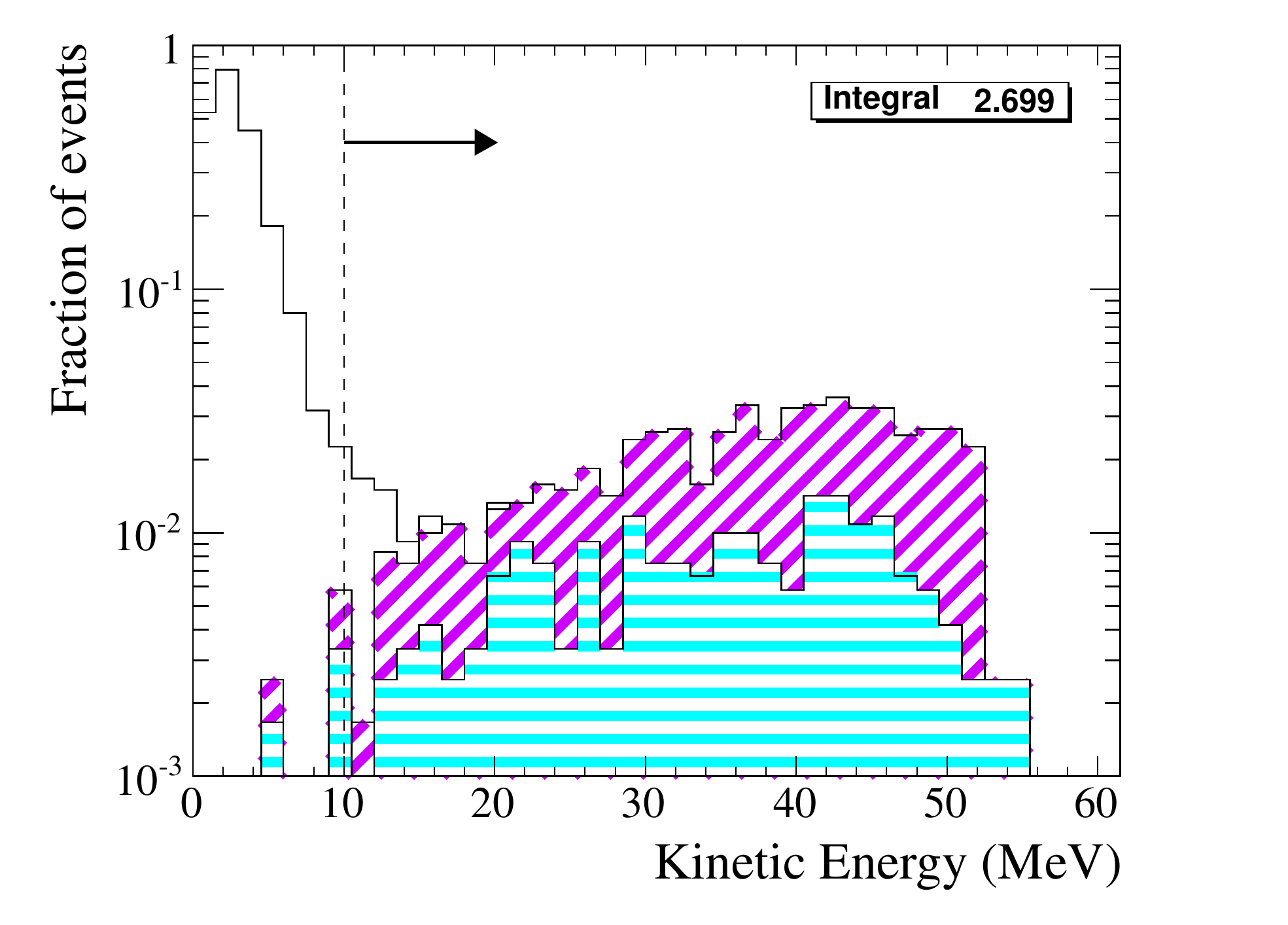} \hfill
\includegraphics[width=0.49\textwidth]{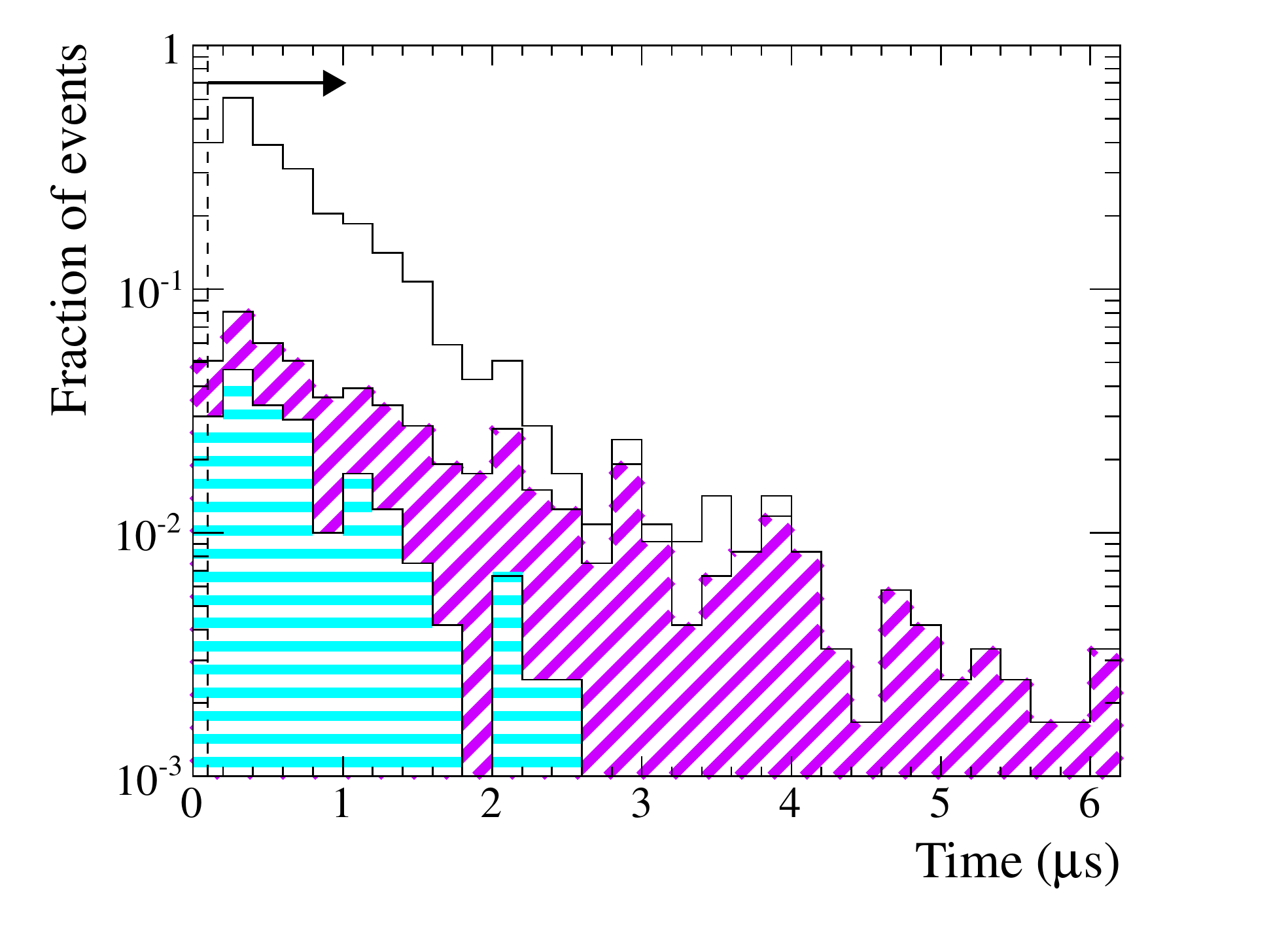} 
\end{center}
\caption{Kinetic energy (left panel) and initial time (right) for particles produced in muon decay or capture processes, originated from the interactions of 4 GeV muon neutrinos in liquid argon. The histogram color and hatching convention is the same as in Fig.~\protect\ref{fig:numubar_decayenergytime}, as is the meaning of vertical lines and arrows.}\label{fig:numu_decayenergytime}
\end{figure}

The kinetic energy and time distributions for muon decay or capture particle products are shown in Figs.~\ref{fig:numubar_decayenergytime} and \ref{fig:numu_decayenergytime}, for 4 GeV muon antineutrino and neutrino interactions, respectively. For muon antineutrino interactions (Fig.~\ref{fig:numubar_decayenergytime}), these distributions are dominated by positrons from $\mu^+$ decays, featuring the characteristic Michel electron energy spectrum and an exponential time distribution compatible with the $\mu^+$ lifetime. For muon neutrino interactions (Fig.~\ref{fig:numu_decayenergytime}), positrons, electrons as well as other low-energy particles ($<$ 10 MeV kinetic energy, mostly neutrons and nuclear de-excitation gammas) are expected. Decay electrons and other particles tend to be produced before decay positrons, because of the competing $\mu^-$ capture process appearing in Eq.~\ref{eq:muminusprocess}.


\subsection{Charge identification in muon neutrino and antineutrino interactions} \label{subsec:MuonCharge_TagEfficiency}

\begin{figure}[t!b!]
\begin{center}
\includegraphics[width=0.60\textwidth]{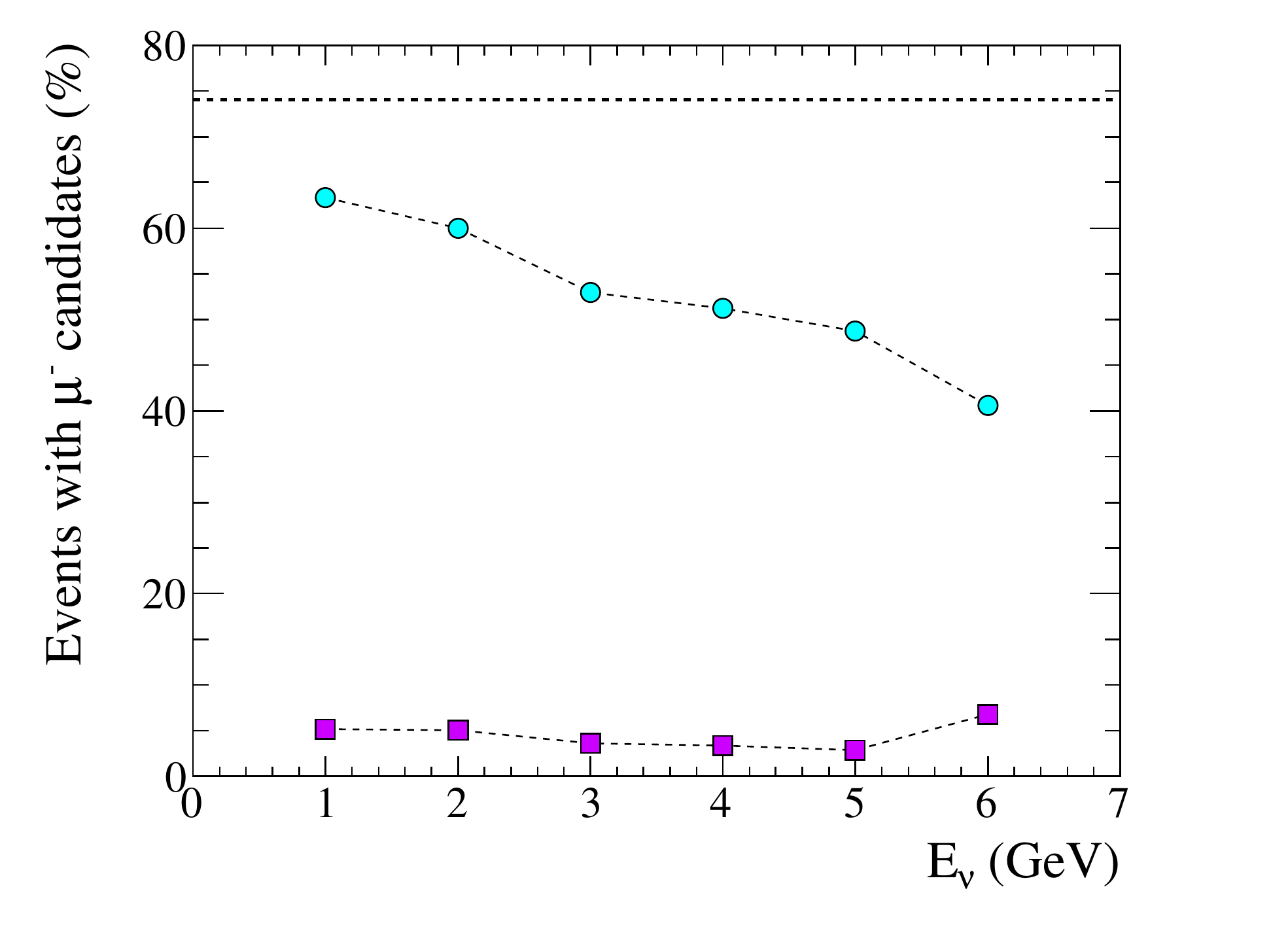} 
\end{center}
\caption{Fraction of events with $\mu^-$ candidates in muon neutrino (cyan circles) and muon antineutrino (violet squares) CC interactions in liquid argon as a function of neutrino energy \Enu, where $\mu^-$ candidates are obtained via the Michel veto requirement (see text for details). The horizontal line at 74\% indicates the $\mu^-$ capture fraction in argon.}\label{fig:muonideff}
\end{figure}

As discussed in Sec.~\ref{subsec:Energy_SecondaryNus}, the scintillation light signal produced in LAr detectors can be used to tag Michel electrons from muon decays. Based on the energy spectra shown in Figs.~\ref{fig:numubar_decayenergytime} and \ref{fig:numu_decayenergytime}, in the following we define a Michel electron candidate as a particle with $>$10 MeV kinetic energy that is produced with some delay with respect to the neutrino interaction. We assume that a minimum delay of 100 ns is necessary to tag a Michel electron from its delayed scintillation light signal. Earlier decays are assumed to be undetectable because of the intense prompt light from the primary neutrino interaction (see Fig.~\ref{fig:ScintTimeDependence}). We further define $\mu^-$ candidate events in muon neutrino or antineutrino CC interactions as those events where no Michel electron candidate is found. The expected fraction of events with $\mu^-$ candidates in muon neutrino and antineutrino CC interactions as a function of neutrino energy is shown in Fig.~\ref{fig:muonideff}. 

For an ideal LAr detector at low energies, we expect a $\mu^-$ identification efficiency of 74\% from a Michel veto requirement, because of the $\mu^-$ capture fraction in argon. In practice, already at 1 GeV neutrino energy a significant amount of muon neutrino interactions produce $\pi^+$'s, in turn producing decay positrons that fail the veto requirement. As a result, the identification efficiency is lower than 74\%, and mildly decreasing with neutrino energy as $\pi^+$ production increases.

Ideally, one would expect muon antineutrino CC interactions to have zero $\mu^-$ misidentification fraction, from a 100\% probability of $\mu^+$ decay. In practice, the 100 ns dead time introduced above to ensure Michel electron tagging causes a misidentification rate of about 4\% at all neutrino energies. In addition, the misidentification rate is expected to increase for increasing neutrino energies, as the fraction of muons escaping from the detector active volume becomes larger. In this work, all muon neutrino interactions are simulated about 20~m upstream of the LAr volume downstream boundary (see Sec.~\ref{sec:Simulation}). This distance corresponds to the range in LAr for muons of about 5 GeV energy. For this reason, the increase in the misidentification rate due to exiting muons in Fig.~\ref{fig:muonideff} can only be seen for $>5$ GeV neutrinos. In practice, this muon charge identification method can only be applied to fully contained final state muons, and the efficiency losses at high energies due to this containment requirement would have to be taken into account.

In summary, and from Fig.~\ref{fig:muonideff}, our simulations and our simple muon capture analysis indicate a 51\% $\mu^-$ identification efficiency for 4 GeV muon neutrino interactions, to be compared with a 3.4\% $\mu^+$ misidentification rate for muon antineutrino interactions of the same energy. Therefore it appears possible to select a statistically significant, highly pure, sample of ``wrong-sign'' muon neutrino interactions in conventional beams operated in antineutrino mode. Such an event sample would provide important input to understanding the beam properties in neutrino oscillation experiments searching for leptonic CP violation via $\nu_{\mu}\to\nu_e$ oscillations. On the other hand, incorrect charge assignment at the few percent level (as the one obtained here) may be too high to efficiently suppress backgrounds in proposed $\nu_{e}\to\nu_{\mu}$ oscillation searches with advanced neutrino beams based on muon storage rings, see \cite{Choubey:2011zzq}. See, however, \cite{Huber:2008yx} for studies on sensitivity to leptonic CP violation and neutrino mass hierarchy with detectors with limited neutrino/antineutrino separation.

\section{Efficiency and feasibility considerations} \label{sec:Feasibility}

Opportunities offered by LAr detectors with enhanced sensitivity to scintillation light have been discussed in Secs.~\ref{sec:Energy} and \ref{sec:MuonCharge}. In the following, we first quantify what is the required light detection efficiency for improved detector performance (Sec.~\ref{subsec:FeasibilityEfficiency}). We then discuss the main experimental ingredients and possible solutions to achieve a sufficiently high scintillation light collection efficiency, arguing that large LAr neutrino detectors fulfilling this goal appear feasible (Sec.~\ref{subsec:FeasibilityImplementation}).

\subsection{Requirements on light detection efficiency} \label{subsec:FeasibilityEfficiency}

In Sec.~\ref{subsec:Energy_Recombination}, we have described how the charge and light signals can be combined to suppress electron-ion recombination effects. For this method to be effective, we require that the Poisson fluctuations in the detected light signal from 1 GeV neutrino interactions do not exceed 1\%. Considering that $\Ngamma\simeq 1.6\cdot 10^7$ scintillation photons are produced in a fully contained 1 GeV neutrino interaction in a 500 V/cm field (see Sec.~\ref{sec:Energy}), we estimate that the fluctuations in the corresponding detected signal are $2.5\cdot 10^{-4}/\sqrt{\varepsilon}$, where $\varepsilon$ is the light collection efficiency. Therefore, the Poisson fluctuations are expected not to exceed the 1\% level for detection efficiencies $\varepsilon > 6\cdot 10^{-4}$.

In Secs.~\ref{subsec:Energy_SecondaryNus} and \ref{sec:MuonCharge}, we have described how the tagging and energy measurement of Michel electrons via their scintillation light signal can improve the neutrino energy reconstruction and neutrino/antineutrino separation of a LAr detector. In order to have a reliable Michel electron detection, we impose two conditions. First, we require that at least 20 photoelectrons (PEs) be detected, from the prompt light fraction alone (assumed to be 0.23 of the total light) for a 10 MeV Michel electron. Second, we require that this signal has a significance of at least 3~$\sigma$ when compared to the fluctuations in the light signal produced by a 4 GeV neutrino interaction. For the latter condition, we conservatively consider the fluctuations in a 10 ns long interval occurring 100 ns after the neutrino interaction, that is after the shortest delayed coincidence considered here. We note that a 100 ns delay is a sufficiently long time for the prompt light from the neutrino interaction to have entirely decayed away, and only the late light from the primary neutrino interaction plays a role, see Fig.~\ref{fig:ScintTimeDependence}. In a 500 V/cm field, the number of detected prompt photons from a 10 MeV electron is about $3.7\cdot 10^4~\varepsilon$, while the number of detected late photons from a 4 GeV neutrino interaction in a $\lbrack 0.10,0.11\rbrack$ $\mu$s interval after the neutrino interaction is about $2.9\cdot 10^5~\varepsilon$. Of the two conditions defined for a reliable Michel electron detection, the second one imposes the most stringent requirement on $\varepsilon$. We estimate that a light collection efficiency $\varepsilon > 1.9\cdot 10^{-3}$ is needed to meet the 3~$\sigma$ detection condition given. In a real detector, the late light may be suppressed because of impurities in argon, and this efficiency requirement may be relaxed somewhat.       

In summary, LAr detectors with scintillation light collection efficiencies in the $10^{-3}$ range would be necessary to improve the calorimetric neutrino energy reconstruction and to provide neutrino/antineutrino separation. 

\subsection{Feasibility considerations} \label{subsec:FeasibilityImplementation}

\begin{table}
\begin{center}
\begin{tabular}{|c|c|c|c|c|c|} \hline
Experiment  & Fiducial    & Coverage      & Reflecting  & Light yield       & Detection \\
            & mass (tons) & (\%)           & surfaces   & (PEs/MeV)         & efficiency       \\ \hline
ICARUS T600 & 476         & 0.5  & No         & $\sim$1 at 0.5 kV/cm \cite{Antonello:2012hg} & $\sim 6\cdot 10^{-5}$  \\ 
LBNO-DEMO   & 300         & 0.5  & No         & $\sim$1 at 0.5 kV/cm \cite{DeBonis:1692375}  & $\sim 6\cdot 10^{-5}$  \\
MicroBooNE  & 70          & 0.9  & No         & $\sim$2 at 0.5 kV/cm \cite{Briese:2013wua}   & $\sim 1.3\cdot 10^{-4}$   \\
CAPTAIN     & 5           & 0.5  & No         & 2.2 at 0.5 kV/cm \cite{Rielage:2013mea}              & $1.4\cdot 10^{-4}$ \\
LArIAT-1    & 0.24        & 0.3  & Yes        & $5\cdot 10^1$ at 0 kV/cm \cite{Szelc:2013ooa}          & $9.8\cdot 10^{-4}$ \\  \hline
ArDM-1t     & 0.85        & 18             & Yes        & $2\cdot 10^3$ at 0 kV/cm \cite{Badertscher:2013soa} & $4\cdot 10^{-2}$ \\ 
DEAP-3600   & 1           & 75             & Yes        & $8\cdot 10^3$ at 0 kV/cm \cite{Boulay:2012hq} & $1.6\cdot 10^{-1}$ \\
DarkSide-10 & 0.01        & 22             & Yes        & $9\cdot 10^3$ at 0 kV/cm \cite{Akimov:2012vv}  & $1.8\cdot 10^{-1}$ \\
 \hline
\end{tabular}
\end{center}
\caption{Scintillation light yield (in PEs per MeV) and light collection efficiency for current LAr detectors used in neutrino physics and for dark matter direct searches, ordered according to increasing efficiency. Information on the detector fiducial mass, the active area coverage of photon detectors, and on the use of reflecting surfaces is also given.}\label{tab:LArLightYield}
\end{table}

The scintillation detection efficiency requirement in Sec.~\ref{subsec:FeasibilityEfficiency} can be contrasted with the efficiency values of current and near-future LAr detectors used in neutrino physics and for dark matter direct searches, see Tab.~\ref{tab:LArLightYield}.  In the table, the efficiencies are typically estimated from the light yields quoted by the experimental collaborations, also given in Tab.~\ref{tab:LArLightYield}, assuming a scintillation light production of $1.6\cdot 10^4$ and $5.1\cdot 10^4$ per MeV at 500 V/cm and zero field, respectively (see Sec.~\ref{sec:Energy}). More accurate measurements exist for current dark matter experiments, where the light signal plays a more important role, while only rough estimates are given for LAr neutrino detectors. 

In current-generation neutrino detectors such as ICARUS T600 \cite{Amerio:2004ze} or MicroBooNE \cite{Chen:2007ae}, scintillation light is read via sparsely populated, large-area cryogenic photomultiplier tubes (PMTs) surrounding the TPC active volume. In order to match the wavelength sensitivity of standard PMT photo-cathodes, the argon scintillation light (128~nm) is shifted to blue light (about 430 nm) by coating either PMTs, or acrylic plates placed in front of them, with tetraphenyl-butadiene (TPB). From Tab.~\ref{tab:LArLightYield}, we can see that improvements in light collection efficiency of about one order of magnitude with respect to these detectors would be needed to reach efficiencies of order $10^{-3}$.

The most straightforward way to obtain enhanced sensitivity to scintillation light in large LAr detectors relies on the use of reflecting surfaces, on a significantly higher active area coverage, or on a combination of both. Information on PMT coverage and on the use of reflecting surfaces is also reported in Tab.~\ref{tab:LArLightYield}. In LAr dark matter detectors, the TPC active volume is surrounded by reflecting panels or foils (typically made of PTFE), coated with a wavelength shifting material (TPB) for increased reflectivity. As a result, detection efficiencies in excess of $10^{-1}$ have been accomplished, although on a small (ton-scale) detector size. In contrast, no reflecting surfaces are typically used in LAr neutrino detectors. One exception is the LArIAT test beam experiment \cite{Adamson:2013/02/28tla}, specifically designed to test the importance of scintillation light information in event reconstruction in LAr TPCs for neutrino physics. In this case, light detection efficiencies of order $10^{-3}$ are expected even with a reduced PMT coverage, thanks to light reflectors. The LArIAT expected light yield has been recently confirmed through a direct measurement on a small scale prototype \cite{Cavanna:2014iqa}. The main drawback of adding reflector panels is the degraded performance to image tracks in LAr using light information. In LAr detectors exposed to pulsed neutrino beams and operated on the surface, the latter capability is useful to identify cosmic ray tracks occurring in coincidence with the beam \cite{Baptista:2012bf}. 

Concerning the active area of photon collection systems, detectors for neutrino physics have typically PMT coverages below 1\%, while coverages in excess of 10\% are common for dark matter detectors, see Tab.~\ref{tab:LArLightYield}. For increased coverage, the simplest solution is to substantially increase the number of large-area PMTs. This approach is viable for current-generation LAr neutrino detectors, which employ only tens of PMTs \cite{Amerio:2004ze,Chen:2007ae}. For future, multi-kton, detectors, less bulky and more cost-effective alternatives for light collection systems may be needed. Several LAr detector R\&D efforts on light detection systems, most notably in the context of the LBNE experiment, are underway and may ultimately provide such solutions. The current plan for LBNE \cite{Bromberg:2013fla} employs wavelength shifting (WLS) adiabatic light guides, as first proposed in \cite{Bugel:2011xg}. This solution uses bars made of cast acrylic, either coated \cite{Baptista:2012bf,Baptista:2013gna} or doped \cite{Gehman:2013kra} with TPB. The bars are polished to provide good capture and transmission characteristics via total internal reflection. Attenuation lengths in excess of 2 meters have been obtained \cite{Bromberg:2013fla}. The bars are read at one end by silicon photomultipliers (SiPMs) or PMTs. While the PMT cost per active area is still lower, SiPMs are becoming more cost-effective. They also offer additional advantages, such as higher photon detection efficiencies and low-voltage power supplies. Alternative photon detection solutions based on WLS fibers coupled to SiPMs are also being explored for LBNE \cite{Wasserman:2013moa}. 

For very large LAr detectors, with linear dimensions of order 10~m or more, special attention to scintillation light attenuation due to impurity contaminations in LAr should also be given. Relatively loose LAr purity specifications are required in this respect even for very large detectors, with a 2~ppm concentration of N$_2$ yielding a measured light absorption length of about 30~m \cite{Jones:2013bca}.

Finally, the requirements on the readout electronics for LAr detectors with enhanced sensitivity to scintillation light are fairly standard, and do not pose significant issues either. Relatively fast signals are needed to distinguish the prompt scintillation light from muon decay electrons from the late light from neutrino interactions. Signal shaping times and data acquisition sampling time intervals of order $\mathcal{O}$(10~ns), and continuous readout over $\mathcal{O}$(10~$\mu$s) around the beam window (that is, $\mathcal{O}$($10^3$) digitized samples) are needed to this end. In addition, the scintillation light readout has to efficiently tag cosmic ray interactions occurring during the entire ionization drift time in the TPCs, which can span several milliseconds in large LAr detectors. In the latter case, a few contiguous digitized samples per cosmic ray interaction at the above indicated $\mathcal{O}$(100~MHz) sampling rate are sufficient. Such a scintillation light readout structure has been adopted, for example, by the MicroBooNE Collaboration, see \cite{Kaleko:2013eda}.

\section{Conclusions} \label{sec:Conclusions}

Neutrino experiments at the multi-kton detector scale and based on the LAr TPC technology are currently being planned to address fundamental questions in neutrino physics. The detector performance of existing LAr neutrino detectors such as ICARUS T600 \cite{Rubbia:2011ft} and MicroBooNE \cite{Chen:2007ae} is largely driven by the information gathered from the ionization signal. Scintillation light from argon de-excitation and electron-ion recombination is also used, as trigger for neutrino interactions, as veto against cosmic rays, and to provide absolute event timing.

In this paper, we have studied additional applications offered by the scintillation light information, for LAr detectors with enhanced light collection efficiencies. We have focused on two key detector performance indicators for neutrino oscillation experiments, namely the neutrino energy reconstruction and the separation between neutrino and antineutrino interactions. Our results are based on the LArSoft simulation framework \cite{Church:2013hea}. Neutrino interactions are modelled according to the GENIE event generator \cite{Andreopoulos:2009rq}, and the charge and light response of a large LAr ideal detector are simulated according to Geant4 \cite{Agostinelli:2002hh,Allison:2006ve} and NEST \cite{Szydagis:2011tk}.

The neutrino energy reconstruction performance has been studied for $\nu_e$ charged-current (CC) interactions in the 1--6 GeV range. We find that scintillation information can help improving the calorimetric neutrino energy reconstruction in two ways. First, the combination of charge and light information can suppress fluctuations in the unavoidable recombination process affecting ionization electrons, given that every recombined electron is expected to produce an additional scintillation photon. In this way, a more compensating LAr calorimeter can in principle be obtained, responding in a more similar way to particles depositing different ionization densities. Second, the scintillation light offers a way to infer the missing energy in the form of secondary neutrinos produced by the decays of muons and pions, by measuring the delayed coincidences from Michel electrons. By combining both improvements, a neutrino energy resolution as good as 3.3\% RMS for 4 GeV $\nu_e$ CC interactions appears possible. This value can be considered as the best resolution achievable for a detector of this type, as it depends on esentially irreducible nuclear effects affecting neutrino interactions and their final state particles, and charge and light quenching effects for energy deposited by nuclei.

Scintillation light can also provide a means to separate muon neutrino from muon antineutrino interactions in a non-magnetized LAr detector. Negatively-charged muons have a high capture probability, about 74\%, on argon nuclei, while this process is absent for positively-charged muons. The presence or absence of a Michel electron candidate in the event, as estimated from scintillation information, can therefore be used to identify muon antineutrino and muon neutrino CC interactions, respectively. By accounting for $\pi^{\pm}$ production in neutrino interactions, for hadronic, decay and capture processes in LAr, and from simple light detection efficiency arguments, we obtain muon neutrino identification efficiencies of about 50\%, and muon antineutrino misidentification rates at the few percent level, for events with fully contained muons.

We argue that the construction of large LAr detectors with sufficiently high light collection efficiencies for the purposes described above appears feasible. Such detectors would require scintillation light detection systems with efficiencies of about $10^{-3}$. While this implies an efficiency increase of about one order of magnitude over existing LAr neutrino detectors, light collection efficiencies as high as $\mathcal{O}$(10\%) have been built for LAr detectors searching for dark matter. In the neutrino physics context, several R\&D activities are now focusing on the development of efficient and cost-effective solutions for light collection systems for 1--10 kton-scale LAr detectors.


\acknowledgments
I am grateful to Janet Conrad, Juan Jos\'e G\'omez Cadenas, Pilar Hern\'andez, Justo Mart\'in-Albo and Brian Rebel for reading early versions of this manuscript, and for suggesting many improvements. This work was supported by the Spanish Ministerio de Econom\'ia y Competitividad under grants CONSOLIDER-Ingenio 2010 CSD2008-0037 (CUP), FIS2012-37947-C04-01 and FPA2011-29823-C02-01, and by the European Community through the FP7 Design Study LAGUNA-LBNO (Project Number 284518).

\bibliographystyle{JHEP}
\bibliography{SciLAr.bib}

\end{document}